\documentclass[aps,twocolumn,pra,notitlepage,]{revtex4-2}
\usepackage{amsmath,amssymb}
\usepackage{dsfont}
\usepackage{graphicx}
\usepackage{physics}
\usepackage{hyperref}
\usepackage{mathtools}
\usepackage{cancel}
\usepackage{bbold}
\usepackage{bm}
\usepackage{color} 
\usepackage[dvipsnames]{xcolor}
\usepackage[T1]{fontenc}

\usepackage{tcolorbox}

\usepackage{soul}

\newcommand{\adiab}[1]{{\tilde #1}}

\usepackage{ulem} 

\begin{document}
\newcounter{theo}
\author{Jan A. Krzywda}\email{j.a.krzywda@liacs.leidenuniv.nl}
\affiliation{$\langle aQa^L
\rangle$ Applied Quantum Algorithms, Lorentz Institute and Leiden Institute of Advanced Computer Science,
Leiden University, P.O. Box 9506, 2300 RA Leiden, The Netherlands}

\author{Łukasz Cywiński}
\affiliation{Institute of Physics, Polish Academy of Sciences, al.~Lotnik{\'o}w 32/46, PL 02-668 Warsaw, Poland}

\title{Decoherence of electron spin qubit during transfer between two semiconductor quantum dots at low magnetic fields}

\begin{abstract}
Electron shuttling is one of the current avenues being pursued to scale semiconductor quantum dot-based spin qubits. Adiabatic spin qubit transfer along a chain of tunnel-coupled quantum dots is one of the possible schemes. In this scheme, we theoretically analyze the dephasing of a spin qubit that is adiabatically transferred between two tunnel-coupled quantum dots. We focus on the regime where the Zeeman splitting is lower than the tunnel coupling, such that interdot tunneling with spin flip is absent. We analyze the sources of errors in spin-coherent electron transfer for Si- and GaAs-based quantum dots. In addition to the obvious effect of fluctuations in spin splitting within each dot, leading to finite \( T_{2}^{*} \) for the stationary spin qubit, we consider the effects activated by detuning sweeps: failure of charge transfer due to charge noise and phonons, spin relaxation due to the enhancement of spin-orbit mixing at the tunnel-induced anticrossing of states localized in the two dots, 
and spin dephasing caused by low- and high-frequency noise coupling to the electron's charge. We show that the latter effect is activated by differences in Zeeman splittings between the two dots. Importantly, all the error mechanisms are more dangerous at low tunnel couplings.  
Our results indicate that away from micromagnets, maximizing the fidelity of coherent transfer aligns with minimizing charge transfer error that was previously considered in J.~A.~Krzywda and \L.~Cywi{\'n}ski, Phys.~Rev.~B {\bf 104} 075439 (2021). For silicon, we suggest having tunnel coupling fulfilling  \( 2t_c \gtrsim 60 \, \mu\)eV when one aims to coherently transfer a spin qubit across a $\sim \!10$ $\mu$m long array of $\sim \! 100$ quantum dots with error less than $10^{-3}$.
\end{abstract}
\maketitle

\section{Introduction}
Coherent coupling of qubit registers separated by a few microns is necessary for the development of scalable quantum computers based on semiconductor quantum dots \cite{Vandersypen_NPJQI17,Boter21,kunne2023spinbus}. Moving a spin qubit across such a distance, i.e., coherent shuttling of a single electron or hole spin, is one of the possible solutions.
Such shuttling can be realized by placing the electron inside a moving potential generated using surface acoustic waves \cite{Takada19,Jadot20} or metallic gates \cite{Seidler_NPJQI22,Langrock_PRXQ23,Xue_NC24,Struck_NC24,Volmer_NPJQI24}. The latter approach has recently allowed for long-distance charge shuttling \cite{Xue_NC24} and coherent spin shuttling on distances of a few hundred nanometers \cite{Struck_NC24,Volmer_NPJQI24,DeSmet_arXiv24} in Si/SiGe structures.

\begin{figure*}[tbh]
    \centering
\includegraphics[width=1\textwidth]{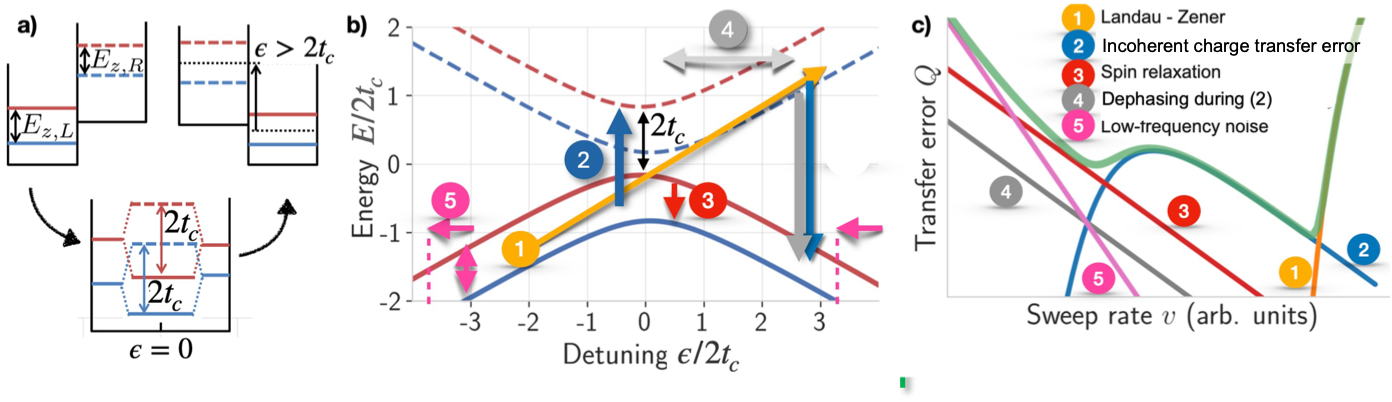}
    \caption{Schematic illustration of the qualitative results of the paper. a) The mechanics of dot-to-dot transition, with the relevant energy scales. The arrows indicate the order of the transition from negative detuning, through avoided crossing at $\epsilon \! = \! 0 $ up to the point where $\epsilon \gg 2t_c$. b) Energy spectrum of the four-level dot-to-dot transition. Orbital ground states are denoted using solid, and orbital excited states are denoted using dashed line for spin-up (red) and spin-down (blue). b) A schematic illustration of the considered mechanisms contributing to coherent transfer error $Q$: (1) the Landau-Zener mechanism of leaving the electron behind in the left dot due to finite sweep rate $v$, (2) incoherent charge transfer error, the process of excitation near the anticrossing due to interaction with environmental noise, followed by relaxation (inelastic tunneling) at large $\epsilon$, (3) spin relaxation enhanced by spin-orbital mixing of states near the anticrossing, (4) spin dephasing due to process (2) activated by difference of Zeeman splittings in the two dots, $\Delta E_Z$, and (5) dephasing caused by low-frequency noise in Zeeman splitting in each dot, and noise in $\epsilon$ when $\Delta E_Z$ is non-zero.
    c) Scaling of these contributions to error $Q$ with the detuning sweep rate $v$, shown in log-log scale.
    At $v$, total error $Q$ is dominated by the infidelity of charge transfer resulting from the interplay between the Landau-Zener process (orange line and Eq.~\ref{eq:LZ}) and incoherent charge transition to higher energy state, followed by possible relaxation (blue line and Eq.~\ref{eq:heal}). At low $v$ the error is dominated by the spin relaxation (red line and Eq.~\eqref{eq:spin_relax}), and loss of coherence between spin eigenstates due to low-frequency noise (violet line and Eqs~\eqref{eq:deps} and \eqref{eq:t2star}), or random time of inelastic tunneling from excited to the ground level (gray line and Eq.~\eqref{eq:phi_transitions}). In the regime of small error $Q \ll 1$ the total error (green line in c)) is a sum of all contributions. }   
    \label{fig:cartoon}
\end{figure*}

Another approach, possible when one-dimensional chains of many tunnel-coupled quantum dots are connecting the registers, relies on sequential adiabatic transfer between neighboring quantum dots \cite{Mills_NC19,Yoneda_NC21,Feng_PRB23, Zwerver_PRXQ23, VanRiggelen_NC24,noiri2022shuttling, DeSmet_arXiv24}: the detuning between two tunnel-coupled quantum dots is changed slowly enough for the electron to move adiabatically from one dot to another, see Fig.~\ref{fig:cartoon}(a) and (b). This method of charge transfer was realized up to four dots in GaAs, in which, however, significant spin dephasing was reported \cite{Fujita_NPJQI17}. More recently, in Si/SiGe quantum dots, a successful transfer of charge across 8 dots \cite{Mills_NC19} was shown, followed by the transfer of a spin eigenstate across a few dots \cite{Zwerver_PRXQ23}, and the transfer of a qubit in a superposition state between two \cite{Feng_PRB23, Yoneda_NC21, noiri2022shuttling,Foster_arXiv24} and five \cite{DeSmet_arXiv24} quantum dots. Coherent shuttling back and forth between two dots has also recently enabled the demonstration of shuttling-based single qubit gates \cite{wang2024operating}.

In this paper, we analyze the sources of error in coherent spin qubit transfer between two tunnel-coupled quantum dots, focusing on Si- and GaAs-based systems. We consider spin relaxation and dephasing caused by electron-phonon interaction, high- and low-frequency charge noise, and hyperfine interaction with nuclei, with special attention devoted to mechanisms that are activated by driving the systems through an anticrossing of energy levels corresponding to electron states localized in the two dots.
We work in the regime of low magnetic fields, defined by requiring the qubit Zeeman splitting, \( E_Z \), to be smaller than the gap at the tunneling-induced anticrossing, \( E_{Z} < 2t_c \), so that there are only two anticrossings (one per spin direction) in the spectrum, see Fig.~\ref{fig:cartoon}(b). For higher \( E_Z \), additional anticrossings between states with opposite spin appear due to spin-orbit interaction, making the spin-flip tunneling coupling finite \cite{Li_PRA17,Zhao_arXiv18,HarveyCollard19PRL,Ginzel_PRB20, Buonascorsi_PRB20,Burkard_RMP23}. As it is very hard to achieve perfectly adiabatic (i.e., without spin-flip) qubit transfer through these anticrossings, interference effects associated with traversing multiple anticrossings are expected to strongly affect both the probability of charge transfer \cite{Krzywda_PRB20} and the spin coherence of the transferred qubit \cite{Ginzel_PRB20}. Optimizing the driving of the two-dot system can counteract these effects \cite{Ginzel_PRB20}, but performing such an optimization for a chain of many QDs is incompatible with the scalability of such an approach to long-distance quantum information transfer \cite{Langrock_PRXQ23}. We thus focus on a more robust case of low magnetic fields, while taking into account the influence of high- and low-frequency noises (phonons and charge noise) on the dynamics of the transferred qubit.

 The fact that some of the error mechanisms discussed here become suppressed with lowering of the \( B \) field provides additional motivation for focusing on low \( B \). For analogous reasons, we assume that the valley splitting, $E_{VS}$, in each of the Si-based quantum dots fulfills \( E_{VS} > 2t_c > E_{Z} \), so that the transfer of an electron initialized in the valley ground state can be treated, to a very good approximation, as if the excited valley state was absent. Note that this requires that the ground-state valleys states in each pair of QDs are in fact appreciably tunnel-coupled. Current understanding of valley coupling in presence of interface disorder indicates that this should be the case not only in the regime of large and deterministic valley coupling, but also in the regime of large disorder-induced randomness in valley coupling \cite{Wuetz_NC22,Losert_PRB23,Losert_arXiv24}.

Let us stress that since the long-term motivation for the research presented here is achieving coherent qubit transfer across approximately 100 quantum dots, we use rather moderate values of tunnel coupling, \( 2t_c < 100 \) \(\mu\)eV. For larger values, coherent dot-to-dot transfer was shown experimentally \cite{Feng_PRB23,Yoneda_NC21,noiri2022shuttling}, but having such strong couplings might be hard to maintain for all pairs of dots in a longer array of them \cite{Langrock_PRXQ23}.

We analyze in detail five mechanisms of errors in coherent spin transfer, which are depicted in Fig.~\ref{fig:cartoon}(b) and (c). We build on the results of previous works \cite{Krzywda_PRB20, Krzywda_PRB21}, where the charge transfer error resulting from a combination of the Landau-Zener mechanism (1) and environmental interactions (2) were considered. In those studies, only a lower bound on the coherent transfer error was provided, since transfer of charge is only necessary, but not sufficient, condition for coherent communication using spin shuttling. 

Here, we explicitly consider the spin degree of freedom, and discuss mechanisms activated by the finite coupling between orbital and spin degrees of freedom, such as spin relaxation \cite{Srinivasa13} (3)  and spin dephasing (4, 5). The latter is triggered by a nonzero difference in Zeeman spin splitting between the dots, \( \Delta E_Z \), due to a magnetic field gradient or differences in the dot \( g \)-factors. The presence of these factors introduces dephasing when  time spent in each dot becomes random \cite{Gawelczyk18,Langrock_PRXQ23}. As we show, this process can occur due to inelastic transfer to an excited state followed by relaxation (mechanism 4 in Fig.~\ref{fig:cartoon}(b,c)), but also during fully adiabatic transfer due to the presence of low-frequency 1/f noise in detuning \cite{Yoneda_NN18,Struck_NPJQI20} (mechanism 5). The dependence of all contributions to the phase error of the shuttled qubit, \( Q \), on the detuning sweep rate, \( v \), is presented schematically in Fig.~\ref{fig:cartoon}(c). One can see that, depending on the relative importance of the contributions, \( Q(v) \) can exhibit nontrivial behavior: there may be two local minima, and both \( Q \propto 1/v \) and \( Q \propto 1/v^2 \) scalings with decreasing \( v \) are possible. It should be noted that while the charge transfer error as function of $v$ exhibited qualitative differences between GaAs and Si, being monotonic in GaAs, and exhibiting local minimum for Si, the total coherent transfer error considered in this paper looks qualitatively similar in the two materials, with very low and large $v$ leading to large errors, although the dominant error mechanism at low $v$ depends on the material and the choice of parameters.

The paper is structured as follows: after using Sec.~\ref{sec:model} to introduce the description of the four-level adiabatic transition and the adiabatic master equation used throughout the paper, in Sec.~\ref{sec:sources}, we briefly introduce the most relevant sources of coherent transfer error: charge transfer error (Sec.~\ref{sec:charge_transfer}), spin relaxation (Sec.~\ref{sec:spin_relax}), and spin dephasing due to charge noise and fluctuating spin splittings (Sec.~\ref{subsec:dephasing}). This is followed by Sec.~\ref{sec:case_study}, in which we compare the predictions for Si- and GaAs-based double quantum dots. 
The paper is concluded by Sec.~\ref{sec:discussion}, in which we discuss the implications of the results for mid-range coherent transfer.

\section{The Model}
\label{sec:model}
\subsection{Electron spin qubit in a double quantum dot}
We consider ground orbital states of two tunnel coupled quantum dots, $\ket{L}$ and $\ket{R}$. In presence of magnetic field, each of them splits into a spin doublet with dot-dependent Zeeman splittings $E_{Z,L}$ and $E_{Z,R}$, see Fig.~\ref{fig:cartoon}(a). Without the spin-orbit interaction the double-dot Hamiltonian is spin-diagonal:
    \begin{equation}
    \label{eq:ham}
        \hat H_{0}(t) = t_c\hat \tau_x + \frac{\epsilon(t)}{2}\hat \tau_z + \frac{E_Z}{2}\hat \sigma_z + \frac{\Delta E_Z}{4} \hat \tau_z \hat \sigma_z.
    \end{equation}
    where $\hat \tau_z = \ketbra{L}{L} - \ketbra{R}{R}$, $\hat \tau_x = \ketbra{L}{R} + \ketbra{R}{L}$, $\Delta E_Z = E_{Z,L}-E_{Z,R}$, $E_Z = \tfrac{1}{2}( E_{Z,L}+E_{Z,R})$, and $\epsilon(t)$ is the time-dependent interdot energy detuning. The eigenenergies can be written as:
    \begin{equation}
    \label{eq:energy_adiab}
        E_{sq}(t) = \frac{\sigma_s}{2}E_Z + \frac{q}{2}\Omega_s(t),
    \end{equation}
    where $\sigma_s = \pm 1$ for spin eigenstates $s=\uparrow,\downarrow$, and $q = \pm \, 1$ for excited and ground orbital states respectively. The spin-dependent orbital gap is given by
    \begin{equation}
    \label{eq:orb_gap}
    \Omega_s(t) = \sqrt{(\epsilon(t) + \sigma_s \Delta E_Z/2)^2 + (2t_c)^2}.
    \end{equation}
    
Next, we add the coupling between spin and orbital degrees of freedom  due synthetic or intrinsic spin-orbit interaction. In the dot basis used above it can be written as \cite{Ginzel_PRB20}:
    \begin{equation}
    \label{eq:Vso}
        \hat V_\text{so} = \frac{b_\perp}{2}\hat \tau_z \hat \sigma_x + \frac{a_{\text{im}}}{2}\hat \tau_y \hat \sigma_x + \frac{a_{r}}{2}\hat \tau_y \hat \sigma_y,
    \end{equation}
    where the $b_\perp$ is caused by the difference of transverse magnetic fields between the QDs due to a magnetic field gradient or a g-tensor difference (synthetic spin-orbit interaction), and remaining terms $a =a_{r} + i\, a_{\text{im}}$ are due to intrinsic spin-orbit coupling. Note that in Si-based quantum dots, presence of electric field in $z$ direction, and atomic roughness at Si/SiGe or Si/SiO$_{2}$ interface activates both Rashba and Dresselhaus spin-orbit terms, so the spin-orbit interaction is qualitatively the same for Si- and GaAs-based QDs \cite{Golub_PRB04,Harvey-Collard_PRL19}- there is only a quantitative difference, with energy scale of this interaction being smaller in Si than in GaAs.

We assume here $2t_c>E_Z$, for which states with opposite spins are always separated in energy. The above spin-orbital coupling introduces then only a weak mixing of the eigenstates  $\ket{\psi_{sq}(t)}$ of Hamiltonian \eqref{eq:ham}, such that the instantaneous eigenstates of total Hamiltonian ${\hat{H}_0(t) + \hat{V}_{\mathrm{so}}}$ can be approximated as:
    \begin{equation}
    \label{eq:pert_states}
        |\psi_{sq}'(t)\rangle = \ket{\psi_{sq}(t)} + \sum_{s'q'} \frac{\bra{\psi_{s'q'}(t)} \hat V_\text{so} \ket{\psi_{sq}(t)}}{E_{sq}(t)-E_{s'q'}(t)} \ket{\psi_{s'q'}(t)},
    \end{equation}
where $\ket{\psi_{sq}(t)}$ are the eigenstates of $\hat H_0(t)$ with energy $E_{sq}(t)$ given in Eq.~\eqref{eq:energy_adiab}. Note that in considered regime of $2t_c>E_Z$, denominator is non-zero for any time instant. This allows us to neglect there modification of the $\hat H_0$ spectrum due to presence of $\hat V_\text{so}$. Since $\bra{\psi_{s'q'}}\hat V_{\text{so}}\ket{\psi_{sq}} \ll 2t_c-E_Z$. The contribution from $\hat V_{\text{so}}$ is relatively small and we use it only to dress the spin-diagonal states  $|\psi_{sq}'(t)\rangle$ in Eq.~\eqref{eq:pert_states}. From now we will drop the prime, and consider all of the states dressed.

\subsection{Interdot transfer of the spin qubit}
We assume the initial spin state is an equal spin superposition in a ground orbital state of the left dot, i.e.~the lowest energy orbital eigenstate of $\hat{H}$ at $-|\epsilon_0| \! \ll \! -|t_c|$, which is well approximated by 
\begin{equation}
   \ket{\Psi(0)} = \frac{|\psi_{\uparrow-}(0)\rangle+ |\psi_{\downarrow-}(0)\rangle}{\sqrt{2}} \approx  \ket{L}\otimes\frac{\ket{\uparrow}+\ket{\downarrow}}{\sqrt{2}} \,\, . \label{eq:Psi0}
\end{equation}
The transfer of the qubit from L to R dot is driven by a linear and symmetric sweep of detuning with characteristic sweep rate $v$, i.e.~$\epsilon(t) = -\Delta\epsilon/2 + vt$, from $t\!=\!0$ to $t\!=\! t_f\!=\! \Delta\epsilon/v$, where $\Delta \epsilon$ is the detuning swing that causes the interdot charge transfer. 

We concentrate on a single figure of merit: the coherence between two lowest in energy instantaneous states at the final time, $t_f$. For $2t_c \! < \! E_Z$ it is simply the spin coherence of an electron in the ground orbital
   \begin{equation}
    W(t) = 2 \langle\psi_{\uparrow-}(t_f)| \hat 
    {\varrho}(t_f)| \psi_{\downarrow-}(t_f)\rangle \,\, , \label{eq:W}
\end{equation}
where $\ket{\psi_{s-}(t_f)}$, with $s = \uparrow, \downarrow$, are two lowest-energy instantaneous eigenstates of $\hat H(t_f)$, previously defined in Eq.~\eqref{eq:pert_states} and $\hat{\rho}(t_f) = \ket{\Psi(t_f)}\bra{\Psi(t_f)}$ (where $\ket{\Psi(t_f)}$ is the state that
evolves from $\ket{\Psi(0)}$ from Eq.~\eqref{eq:Psi0}under influence of $\hat{H}(t)$) is the density matrix of the electron spin in the double quantum dot.
For $\epsilon(t_f) \! \gg \! |t_c|$ the target eigenstates  $\ket{\psi_{s-}(t_f)}$ are well approximated by $\ket{R}\otimes\ket{s}$, i.e.~the Zeeman doublet of states of qubit located in the $R$ dot.

We define the transfer error $Q$ as the probability of all the events that do not lead to a transfer of the electron between two quantum dots with its spin coherence preserved:
\begin{equation}
    Q =  1 - |W(t_f)| \,\, .
\end{equation}
According to this definition, failure of charge transfer (i.e.~occupation of $\ket{R}$ orbital state smaller than $1$ at the end of detuning sweep), spin relaxation, and spin dephasing, all contribute to $Q$. Note that the modulus in the above definition removes the deterministic phase acquired by the spin superposition state during the interdot transfer of the qubit.

For a closed four-level system in regime of low magnetic fields ($E_{Z} \! < \! 2t_c$), the spin dynamics during charge transfer can be neglected, and the only source of transfer error $Q$ is simply the failure of interdot charge transfer. We focus here on the adiabatic charge transfer regime, $v\ll (2t_c)^2$, in which the charge transfer error is exponentially suppressed, as predicted by the Landau-Zener formula \cite{Shevchenko_PR10}:
\begin{equation}
\label{eq:LZ}
    Q_\text{LZ} = \exp(-2\pi\frac{t_c^2}{v}) \,\, ,
\end{equation}
and shown as a yellow line in Fig.~\ref{fig:cartoon}c. Note that we use units in which $\hbar\! =\! 1$.

\subsection{Dynamics in presence of coupling to environment}
The physics of coherent transfer error $Q$ is thus rather trivial for a closed system at low magnetic fields. However, a realistic description of interdot transfer of a spin qubit should take into account the interaction with environment relevant for semiconductor quantum dots: a bath of lattice vibrations (phonons) causing transitions between electronic energy levels \cite{Srinivasa13}, nuclear spins coupling directly to the spin \cite{Chekhovich_NM13}, and an environment containing sources of charge noise: Fermi sea of electrons in nearby metallic gates (giving rise to Johnson-Nyquist noise), and an ensemble of two-level systems with finite electric dipole moments causing the $1/f$ noise \cite{Paladino_RMP14,Kepa_spin_APL23,Connors_NC22,Kepa_charge_APL23}.

Electric field noise leads to fluctuations of mean position of electron localized in each of the dots. In presence of finite gradient of $z$ component of the magnetic field, these lead to fluctuations in Zeeman splittings $E_{Z,L(R)}$ in both QDs \cite{Yoneda_NN18,Struck_NPJQI20}. Similar fluctuations are caused by hyperfine interaction with nuclei in each of the dots \cite{Hanson_RMP07,Cywinski_APPA11,Chekhovich_NM13}. We will include these processes later in the analysis, and now we will focus on processes that are activated by interdot tranfer of the electron caused by detuning sweep. These require the coupling of the environment to the charge states only:
\begin{equation}
\label{eq:ve}
    \hat V_{e} = \frac{\hat V_\epsilon}{2} \hat \tau_z + \frac{\hat V_{t_c}}{2} \hat \tau_x,
\end{equation}
where the $\hat V_\epsilon, \hat V_{t_c}$ operators acts on the environmental degrees of freedom and represent quantum fluctuations in detuning and tunnel coupling respectively. Such environment is typically assumed to be in thermal equilibrium at low, but finite electron temperature of $T\approx 100\,$mK such that $2t_c/k_\text{B}T \gtrapprox  1$ \cite{Vandersypen_NPJQI17}. The high-frequency environmental noise causes inelastic transitions between the instantaneous eigenstates of Eq.~\eqref{eq:ham} , which we model by Adiabatic Master Equation (AME) \cite{albash2012quantum, nalbach2014adiabatic, PhysRevE.95.012136} of the form:
   \begin{align}
   \label{eq:ame}
   \partial_t&{ {\tilde {\varrho}}}(t)= -i\comm{ {\tilde {H}_0}(t) + \partial_t \hat A^\dagger(t) \hat A(t)}{{\tilde \varrho}(t)} +  \\
   &-\sum_{q=\pm}\Gamma_{q,\text{spin}}(t) \bigg(\nonumber{\tilde \sigma}_{q}{\tilde \varrho}(t) {\tilde \sigma}_{q}^\dagger -\frac{1}{2}\acomm{{\tilde \sigma}_{q}^\dagger{\tilde \sigma}_{q}}{{\tilde \varrho}(t)}\bigg) \nonumber \\
   &-\sum_{q=\pm,s=\uparrow\downarrow} \Gamma_{q,\text{charge}}^{(s)}(t) \bigg(\nonumber{\tilde \tau}_{sq}{\tilde \varrho}(t) {\tilde \tau}_{sq}^\dagger -\frac{1}{2}\acomm{{\tilde \tau}_{sq}^\dagger{\tilde \tau}_{sq}}{{\tilde \varrho}(t)}\bigg) \,\, ,
\end{align}
where all of the operators and the states are written in the \textit{adiabatic frame}, ${\tilde O} = \hat A(t)\hat O(t) \hat A^\dagger(t)$, $\ket{\psi(t)} = \hat A(t) |\tilde \psi(t)\rangle$ i.e.~in the basis associated with the instantaneous states of $\hat H_0(t)$ (see Sec.\ref{sec:adiabatic_states} for more details). By definition, the transformation associated with an operator $\hat A(t)$ diagonalizes the Hamiltonian at every time instant:
\begin{equation}
\adiab H_0(t) = \hat A(t) \hat H_0(t) \hat A^\dagger(t) = \sum_{s=\uparrow,\downarrow} 
\frac{1}{2}\bigg(E_Z\sigma_s + \Omega_s(t) \adiab{\tau}_z\bigg) \ketbra{s}.
\end{equation}
Its time-dependence generates the coherent coupling, $\partial_t \hat A^\dagger(t) \hat A(t)$ between the orbital eigenstates, 
which is responsible for non-adiabatic evolution that gives $Q_{\text{LZ}}$ from Eq.~\eqref{eq:LZ}. The remainder of the error $Q$ is caused by the interaction with the environment. 

Firstly, we consider spin-diagonal transitions between the orbital states. After transformation the operators ${\tilde{\tau}}_{s\pm} = |\tilde \psi_{s\pm}(t)\rangle\langle \tilde \psi_{s\mp}(t)|$ are responsible for such transitions between the eigenstate of energy ${E_{s\pm} = \sigma_s E_Z/2\pm\Omega_s/2}$ and the eigenstate with energy ${E_{s\mp} = \sigma_s E_Z/2\mp\Omega_s/2}$, respectively. With this process we associate corresponding time-dependent rates $\Gamma_{\pm, \text{charge}}^{(s)}(t)$ of excitation/relaxation for spin-s. Secondly we add transitions between the eigenstates that are adiabatically linked to the opposite spins, i.e. ${\tilde{\sigma}}_{+} = |\tilde \psi_{\uparrow\pm}(t)\rangle\langle \tilde \psi_{\downarrow\pm}(t)|$, ${\tilde{\sigma}}_{-} = |\tilde \psi_{\downarrow\pm}(t)\rangle \langle \tilde \psi_{\uparrow\pm}(t)|$, with associated time-depedent transition rates $\Gamma_{+,\text{spin}}(t)$, $\Gamma_{-,\text{spin}}(t)$. A more detailed derivation of AME can be found in Appendix.~\ref{app:AME}, while the charge and spin relaxation rates are computed in Appendix.~\ref{app:relax}.

Finally, in most experimentally relevant scenarios, spin shuttling is performed in the presence of a low-frequency (slower than the timescale of a single dot-to-dot charge transfer) fluctuations of electric fields and spin splittings in both dots that translate to  quasistatic noise in the parameters of Hamiltonian $
    \hat H(t) = \hat H_0(t) + \delta \hat H$, where
\begin{equation}
        \delta\hat H = \delta t_c\hat \tau_x + \frac{\delta \epsilon}{2}\hat \tau_z + \frac{1}{2}\bigg(\frac{\delta E_{Z,L}}{2}\ketbra{L} + \frac{\delta E_{Z,R}}{2}\ketbra{R}\bigg)\hat \sigma_z.
    \end{equation}
We compute their influence on the qubit coherence by averaging  results of Eq.~\eqref{eq:ame} over quasistatic and independent fluctuations of detuning $\delta \epsilon$, tunnel coupling $\delta t_c$, and spin-splittings in the left $\delta E_{Z,L}$ and in the right dot $\delta E_{Z,R}$. Together, the total contribution to the error from both high-frequency and low-frequency noise can be written as:
\begin{equation}
    Q = 1 - 2 \Big|\overline{\langle \tilde \psi_{\uparrow-}(t_f)|{\tilde  \varrho}(t_f)|\tilde \psi_{\downarrow-}(t_f)\rangle}\Big|,
\end{equation}
where the horizontal line inside the absolute value, represents classical averaging over distribution of $\delta t_c$, $\delta \epsilon$ and $\delta E_{Z,L/R}$.

The numerical simulations of averaged AME are performed by adapting the method of solving the time-dependent Master Equation, available in the QuTiP Quantum Toolbox \cite{johansson2012qutip}. The source code used to generate the figures in the manuscript can be found in a dedicated GitHub repository \cite{github}.

\subsection{Parameters for GaAs and Si quantum dots}
\label{app:parameters}

\begin{table}[htb!]

\begin{tabular}{|c|cc|}
\hline
\textbf{Quantity}                          & \multicolumn{1}{c|}{\textbf{Si}} & \textbf{GaAs } \\ \hline \hline
Tunnel coupling $2t_c$                      & \multicolumn{2}{c|}{40,\,100 $\mu$eV}                          \\ \hline 
Range of detuning sweep $\Delta \epsilon$           & \multicolumn{2}{c|}{0.5 meV}                                         \\ \hline
Temperature $T$                            & \multicolumn{2}{c|}{$100$ mK}                                      \\ \hline
Spin splitting $E_Z$                       & \multicolumn{2}{c|}{$30\,\mu$eV}                                    \\ \hline
Spin splitting difference $\Delta E_Z$    & \multicolumn{1}{c|}{$0.001,1\,\mu$eV }  & $0.1\,\mu$eV                           \\ \hline
Charge relaxation rate $\Gamma_-(2t_c)$    & \multicolumn{1}{c|}{$0.1\,\text{ns}^{-1}$}   & $10\,\text{ns}^{-1}$                         \\ \hline
Dephasing time $T_2^*$     & \multicolumn{1}{c|}{20 $\mu$s}    & 2 $\mu$s                         \\ \hline
Intrinsic SOC $|a|$         & \multicolumn{1}{c|}{$0.1\,\mu$eV}  & $1\,\mu$eV                        \\ \hline
Synthetic SOC $b_\perp$ & \multicolumn{2}{c|}{$0,\,1\, \mu$eV}   \\
\hline
\end{tabular}

\caption{Parameters used in the paper for isotopically purified Si devices and GaAs spin qubit with nuclear spin noise estimation}
\label{tab:params}
\end{table}

In this paper, we take into consideration two contrasting and experimentally relevant cases of isotopically purified Si and GaAs DQD devices. They represent qualitatively different regimes in terms of charge relaxation rate (see analysis in \cite{Krzywda_PRB21}), but also in terms of interaction between the environment and spin degree of freedom analysed in this paper. For instance, due to relatively weaker spin-orbit interaction and the limited number of nuclear spins, isotopically purified silicon is considered a more promising candidate for coherent spin transfer.

For both of the material platforms the charge relaxation rates $\Gamma_{-,\text{charge}}^{(s)}(t)$, $\Gamma_{-,\text{spin}}(t) $ are computed in Appendix~ \ref{app:relax_params}. In the polar GaAs the dominant role in the charge transfer is played by relatively strong coupling to piezoelectric phonons, which can improve charge transfer at low sweep rates \cite{Krzywda_PRB21}. However due to inevitable presence of nuclear spins, coherent transfer that takes longer than $T_2^*\approx 10$ns is impossible without additional control. For this reason we consider here GaAs device with a narrowed nuclear field \cite{Klauser_PRB06,Bluhm_PRL10,Shulman_NC14}. In such a case the effective dephasing time can be extended by a few orders of magnitude, i.e. $T_{2,\text{GaAs}}^*\approx 2\mu$s, at the cost of continuous estimation of the nuclear field \cite{berritta2024real}. We compare such a case against isotopically purified Si (with 800 ppm), with an order of magnitude longer $T_{2,\text{Si}}^* = 20\mu$s \cite{Wuetz_NC23}. 

We assume the detuning and tunnel coupling are affected by the charge noise of 1/f-type with a value of spectral density at $f= 1$Hz given by $S(1\text{Hz})\approx (0.5)^2\mu$eV$^2$/Hz \cite{Connors_NC22,Yoneda_NN18}, that translates to standard deviation of $\langle \delta \epsilon^2 \rangle \approx 5\mu$eV. For tunnel coupling noise we assume significantly weaker amplitude $\langle \delta 2t_c \rangle \approx 0.5\mu$eV, which is consistent with a typical difference between the lever arm of plunger and barrier gates \cite{unseld20232d}. We add the second source of charge noise in form of Johnson noise from wiring with $50$ $\Omega$ impedance \cite{Krzywda_PRB21, Huang_PRB13}. For spin-splitting noise we assume fluctuations in the dots $\delta E_{Z,L}$ and $\delta E_{Z,R}$ are independent, and their standard deviation can be related to typically measured $T_2^*$, i.e. $\langle \delta E_{Z,L}^2\rangle = \langle \delta E_{Z,R}^2\rangle = \sqrt{2}/T_{2}^*$. We assume the $T_2^*$ is due the presence of the nuclei, or charge noise affecting the Zeeman splitting in presence of B field gradient \cite{Neumann_JAP15,dumoulin2021low}.

In both devices we use two values of tunnel coupling, $2t_c = 40$ and $100$ $\mu$eV, and set $E_Z = 30\mu$eV corresponding to $B\approx 0.25$T in Si, such that assumed condition $E_Z < 2t_c$ is met. In Si we use two contrasting values of a difference in Zeeman splitting between the dots i.e. $\Delta E_Z = 1,10^{-3}\mu$eV, where the first of them is more typical in presence of intentional gradient \cite{undseth2023nonlinear}, for instance in the manipulation zone of the shuttling-based quantum computer  architecture \cite{kunne2024spinbus}, while the latter can be associated with a typical difference in g-factors $\delta g/g \lesssim 10^{-3}$ \cite{patomaki2024pipeline}. For GaAs DQDs we use rather small $\Delta E_Z = 0.1\mu$eV, but compare intrinsic spin-orbit coupling $a_\text{im} = 1\mu$eV, against more optimistic $a_\text{im} = 0.1\mu$eV. In Si we keep $a_\text{im}=0.1\mu$eV. To avoid interference we assume $a_r=0$. For both devices, we compare intrinsic spin orbit coupling against synthetic spin-orbit coupling $b_\perp$, produced by the longitudinal gradient of magnetic field, for instance in vicinity of manipulation region. We use value of $b_\perp \! = \! 1\, \mu$eV, which in Si corresponds to the gradient of the order of $0.1\,$mT/nm and dots separation $d\approx 100\,$nm.

\section{Sources of coherent transfer error}
\label{sec:sources}
We concentrate here on small errors, i.e. $Q\ll 1$ for which in the leading order we have
\begin{equation}
\label{eq:dwm}
Q \approx Q_\text{charge} + Q_\text{relax} + Q_\varphi \,\, ,
\end{equation}
with the total error separated into contributions due to charge transfer error 
\begin{equation}
Q_\text{charge} = (Q_\uparrow + Q_\downarrow)/2,
\end{equation}
where $Q_s$ is the loss of occupation of s-spin eigenstate, i.e. $Q_{s} = 1 - 2|\langle\tilde \psi_{s-}(t_f)|{\tilde \varrho}(t_f)|\tilde \psi_{s-}(t_f)\rangle|^2$, the spin relaxation \begin{equation}
\label{eq:qrelax}
Q_\text{relax} \approx \frac{1}{2} \int_{0}^{t_f} \Gamma_{-\text{spin}}(t)\text{d}t,
\end{equation}
where $\Gamma_{-,\text{spin}}(t)$ is time-dependent spin relaxation and pure dephasing contribution 
\begin{equation}
Q_\varphi \approx \frac{1}{2}\langle\delta \varphi^2 \rangle,
\end{equation}
where $\langle \delta\varphi^2 \rangle$ is the square of random phase $\delta \varphi$ averaged over realisations of low-frequency noise and random time spent in the excited state. 
Each of them will be analyzed separately in the remainder of  Sec.~\ref{sec:sources} and combined with the parameters corresponding to Si and GaAs devices in Sec.~\ref{sec:case_study}

\subsection{Charge transfer error $Q_\text{charge}$}
\label{sec:charge_transfer}
Before considering spin degree of freedom, for completeness we start by a summary of results from Ref.~\cite{Krzywda_PRB21}, where we have analysed the necessary condition for coherent communication, i.e. the high-fidelity if the  charge transfer. With sufficiently weak spin-orbit interaction and in the relevant regime of $2t_c > E_Z$, the spin components can be treated as uncoupled, and hence in the adiabatic regime (i.e. where $Q_\text{LZ}\ll 1$), the charge transfer error follows the result of \cite{Krzywda_PRB21}:
\begin{equation}
\label{eq:Q_int}
    Q_\text{charge} \approx  \int_{0}^{t_f}\, \text{d}t\,  \Gamma_{+}(\Omega[t]) e^{-\int_{0}^{t} \big(\Gamma_{+}(\Omega[t']) + \Gamma_{-}(\Omega[t']) \big)\text{d}t'}.
\end{equation}
The above equation can be derived from the AME from Eq.~\eqref{eq:ame}), with the unitary part neglected, i.e.~in the limit of $(2t_c)^2 \gg v$. It describes the occupation of the excited state obtained by integrating the excitation rate, which is normalised by the total number of transitions. 
For time-independent rates, the long-time limit would give the equilibrium occupation $Q_\text{eq} = \Gamma_+/(\Gamma_- + \Gamma_+)$. However this prediction is expected to be heavily modified for time-dependent rates \cite{VogelsbergerPRB2006}, for instance if $\Gamma_+$ is active only locally, around avoided crossing. This is often the case in thermal equilibrium, at which excitation and relaxation rates are related by the detailed balance condition $\Gamma_+(\Omega[t]) = \Gamma_-(\Omega[t]) e^{-\beta \Omega[t]}$ \cite{you2021positive}, where $\beta = 1/k_\text{B}T$, so $\Gamma_+ \! \ll \! \Gamma_{-}$ in most of the parameter regime of interest.

\begin{figure}
\centering\includegraphics[width=\columnwidth]{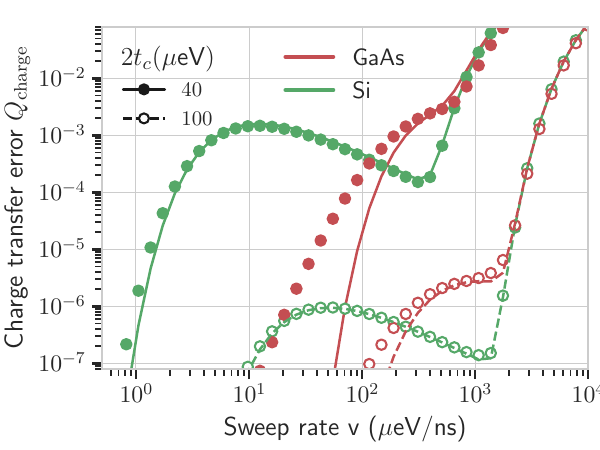}
    \caption{Charge transfer error $Q_\text{charge}$ as a function of sweep rate $v$ for the GaAs (red) and Si (green) DQD devices described in Section~\ref{app:parameters} for two values of tunnel couplings $2t_c = 40\,\mu$eV (filled dots) and $2t_c = \,100\mu$eV (hollow dots). Dots correspond to the numerical solution of Adiabatic Master Equation \eqref{eq:ame}, and lines depict analytic formula for $Q_\text{HEAL}$ \eqref{eq:heal} plus the contribution from Landau-Zener mechanism, i.e. $Q_\text{charge} \approx Q_\text{LZ}+Q_{\text{HEAL}}$. 
    We used the relaxation rates computed in Appendix~\ref{app:relax_params}.}
    \label{fig:charge}
\end{figure}

When coupling to environment is sufficiently weak, i.e. $\Gamma_+(2t_c) t_c/v \ll 1$, one can consider only a single transition from ground to excited state, that is followed by a possible relaxation at finite detuning. This approach, termed Healed Excitation Approximation Limit (HEAL) in Ref.~\cite{Krzywda_PRB21}, resulted in a approximate formula:
\begin{equation}
\label{eq:heal}
    Q_\text{HEAL} \approx \Gamma_{+}(2t_c) \frac{\sqrt{4\pi k_\text{B}T t_c}}{v} e^{-\chi_\text{HEAL}}.
\end{equation}
This result shows two ways of reducing incoherent charge transfer error. For relatively fast sweeps, i.e. for the healing factor \({\chi_\text{HEAL} = \int_0^{t_f}\Gamma_-(t')\text{d}t' \ll 1}\), the error can be reduced by increasing $v$ since $Q\propto 1/v$. This is because fast sweeps reduce the time around avoided crossings and hence limit the excitations. On the other hand, slower sweeps, for which  $\chi_\text{HEAL} > 1$, enable correction of excitations through subsequent relaxation. This implies that $Q_\text{charge}$ can start decreasing below a certain $v$, as electrons in the excited state relax back to the ground state, reaching the target dot through inelastic tunneling, see Fig.~\ref{fig:charge}). 

The subsequent sections will explore the extent to which such slow transfer of charge is accompanied by spin dephasing and spin relaxation.

\subsection{Spin relaxation error $Q_\text{relax}$}
\label{sec:spin_relax}
In presence of finite spin-orbit interaction Eq.~\eqref{eq:Vso}, the instantaneous eigenstates are no longer spin diagonal, and even purely orbital character of environmental coupling (see Eq.~\eqref{eq:ve}), can induce transition between the two lowest instantaneous eigenstates, which define the spin qubit. 

In this limit, the most relevant contribution comes from the mixing between $\ket {\psi_{\downarrow+}(t)}$ and $\ket{\psi_{\uparrow-}(t)}$ states around the avoided crossing. Since $E_Z> k_\text{B}T$ we concentrate on spin relaxation, i.e. inelastic transition between the states $| \psi_{\uparrow-}(t)\rangle$ and $| \psi_{\downarrow-}(t)\rangle$, which has dominant contribution from:
\begin{align}
\label{eq:gam_mix}
   \Gamma_{\text{spin}}(\Omega[t]) \approx \frac{|\langle  \psi_{\uparrow+}(t)|\hat V_\text{so}|\psi_{\downarrow-}(t)\rangle|^2}{(\Omega(t) - E_Z)^2} \Gamma_{-,\text{charge}}(E_Z)
\end{align}
see Appendix \ref{app:relax} for derivation.

\begin{figure}
    \centering
    \includegraphics[width=\columnwidth]{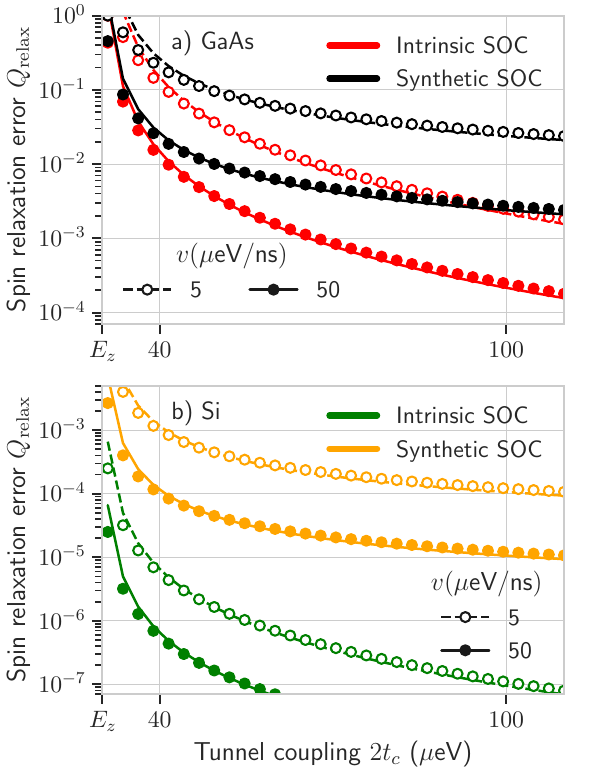}
    \caption{Probability of spin-relaxation as a function of tunnel coupling for two different sweep rates $v = 5,50\,\mu$eV/ns (solid and hollow dots). We compare numerical simulations of Adiabatic Master Equation (dots) against theoretical predictions from Eq.~\eqref{eq:qrelax} (lines) for GaAs (red) and Si (green) devices, that are characterized by the relaxation rates from Appendix~\ref{app:relax_params}. For each device we compare contribution from intrinsic and synthetic spin-orbit coupling with the parameters given in Table~\ref{tab:params}. For the adiabatic master equation, we initialize the system in the excited spin state and assume effectively zero temperature, such that $\Gamma_+ = 0$. We confirm a good agreement between numerical and analytical methods in the relevant region of $2t_c>E_Z = 30\,\mu$eV.
    }
    \label{fig:spin_relax}
\end{figure}

The relaxation rate is upper bounded by its value at the avoided crossing. To compute probability of the spin-flip around avoided crossing $Q_\text{relax}$ from Eq.~\eqref{eq:qrelax}, we assume the relaxation rate is constant and non-negligible only around avoided crossing. Since the electron spends there a time period of the order of $2t_c/v$, we can estimate
\begin{equation}
\label{eq:spin_relax}
 Q_\text{relax} \approx \left|\frac{aE_Z + 2b_\perp t_c}{(2t_c)^2 - E_Z^2}\right|^2 \frac{2t_c}{v} \,\Gamma_{-,\text{charge}}(E_Z).
\end{equation}
Note that while the contribution from intrinsic spin-orbit coupling, $\propto a E_{Z}$, vanishes as $E_Z$ goes to zero due to Van Vleck cancellation \cite{VanVleck40,Hanson_RMP07}, the contribution from the synthetic spin-orbit coupling, $\propto b_\perp t_c$, does not exhibit this effect, as discussed in \cite{Huang_NJP22}.
In Fig.~\ref{fig:spin_relax} we compare the above simple formula against solution of AME in the limit of effectively zero temperature, i.e. when $\Gamma_+ = 0$, such that the dots represents numerically simulated probability of spin relaxation during the transfer. On the x-axis we mark two tunnel couplings considered in this paper $2t_c = 40,100\mu$eV. 

The figure indicates a significantly higher predicted probability of spin-relaxation in GaAs, compared to Si, attributed to larger intrinsic spin-orbit coupling (approximately by an order of magnitude, $\abs{a}= 1\mu$eV) and a much faster relaxation rate at avoided crossings due to piezoelectric coupling. Even outside the manipulation region (red, with $b_\perp = 0$), the resulting $Q_\text{relax}\geq 10^{-3}$ at $v = 5\mu$eV/ns poses a serious threat to coherent transfer in GaAs. Conversely, the intrinsic spin-orbit interaction in Si (green) induces an effectively negligible error, $Q_\text{relax}<10^{-5}$ at $2t_c>40\mu$eV. Furthermore, Eq.~\eqref{eq:spin_relax} indicates that the intrinsic SOC contribution scales as $Q_\text{relax} \propto E_Z^2/t_c^3$, suggesting potential reduction with increased tunnel coupling or decreased magnetic field (Van Vleck cancellation).

However, if the dominant spin-orbit coupling contribution arises from the transverse gradient of the magnetic field $b_\perp$, with $b_\perp = 1\mu$eV as an example corresponding to a gradient of $\sim 0.1$mT/nm and dots separation of $d \approx 100$nm. In such a case the scaling becomes $Q_\text{relax} \propto 1/t_c$. In GaAs (black), this contribution becomes significantly different from the intrinsic case only at large $2t_c$ \cite{VanVleck40}. In Si (orange), it increases the relaxation rate by about two orders of magnitude, resulting in $Q_\text{relax} \approx 10^{-3}$ at $2t_c = 40\,\mu$eV and $v = 5\,\mu$eV/ns, potentially becoming a significant contribution to transfer error. It's important to note that the error is expected to further increase as the tunnel coupling approaches $E_Z$, eventually being affected by interference effects \cite{Krzywda_PRB20}, which are beyond the scope of this paper.

\subsection{Spin dephasing $\langle \delta \varphi^2 \rangle$}
\label{subsec:dephasing} 
As the last source of error during coherent transfer we discuss pure dephasing effects, caused by the uncontrolled, random evolution of the phase of electron qubit during the transition.
This randomness appears when $\Delta E_Z $ is nonzero due to dot-dependent Zeeman splitting, and when the time  spent in each of the dots is a random variable. This randomness is caused by incoherent transitions between instantaneous eigenstates due to presence of both low- and high- frequency fluctuations of electric field, as well as electron-phonon coupling. It can be parameterized by a random excess of time spent in one of the dots, $\delta t$, which translates into stochastic contribution to a relative phase between the spin eigenstates:
\begin{equation}
\label{eq:def_dt}
    \delta \varphi \approx \Delta E_Z \delta t.
\end{equation}
In the fully adiabatic case, and for a symmetric sweep, the electron on average spends the same amount of time in each dot, i.e. $\langle \delta t \rangle = 0$ and dephasing is related to non-zero $\langle \delta t^2 \rangle$. Note that even if a given mechanism leads to finite $\langle \delta t \rangle$, this simply gives renormalization of the deterministic phase acquired by the spin during the shuttling, while the presence of nonzero $\langle \delta t^2 \rangle - \langle \delta t  \rangle^2$ leads to dephasing.
In general, we expect $\Delta E_Z \ll 2t_c$, which allows to treat $\Delta E_Z$ as small perturbation. 

\subsubsection{Dephasing due to high-frequency noise}
As a first relevant mechanism leading to $\langle \delta t^2 \rangle\neq 0$, we consider the dephasing caused by high-frequency-noise induced transitions   between the eigenstates of Fig.~\ref{fig:cartoon} b). They are caused by the energy exchange between the electron and the environment, that is in resonance with the instantaneous orbital gap $\Omega_s(t)$ from Eq.~\eqref{eq:orb_gap}. In the limit of a single excitation at avoided crossing, the contribution $\langle \delta \varphi^2 \rangle$ is due to variance of distribution of time spent in the excited orbital state. This time period is set by the time between the excitation and subsequent relaxation, which brings the electron charge back to the ground state. To illustrate this mechanism, we set the excitation probability to $P_e$ and the relaxation rate to a constant $\Gamma_-$. In such a case the waiting time is given by the distribution $p(\delta t) = \Gamma_- e^{-\Gamma_- \delta t}$, with a variance $\langle \delta t^2 \rangle = \Gamma_-^{-2}$. Directly from the above we have \cite{Gawelczyk18}:
\begin{equation}
\label{eq:phi_transitions}
    \langle \delta \varphi^2\rangle_\text{non-adiab} = P_{e} \left(\frac{\Delta E_Z}{\Gamma_-}\right)^2,
\end{equation}
where $P_e$ is the probability of excitation from ground to excited state around avoided crossing. In absence of coherent excitations (L-Z), it can be estimated using Eq.~(\ref{eq:heal}) with $\chi_\text{HEAL} = 0$ (before the relaxation), such that $P_e\approx \Gamma_{+}(2t_c) \sqrt{4\pi k_\text{B}T t_c}/v$ \cite{Krzywda_PRB21}. This results in phase error that scales as
\begin{equation}
\langle \delta \varphi^2\rangle_\text{non-adiab} \propto \sqrt{2t_ck_BT}\frac{(\Delta E_Z)^2}{v \Gamma_-(2t_c)} e^{-\beta 2 t_c} \label{eq:Qnonadiab_scaling}
\end{equation}
with the two most easily tunable parameters.

\begin{figure}[htb!]
    \centering
    \includegraphics[width=\columnwidth]{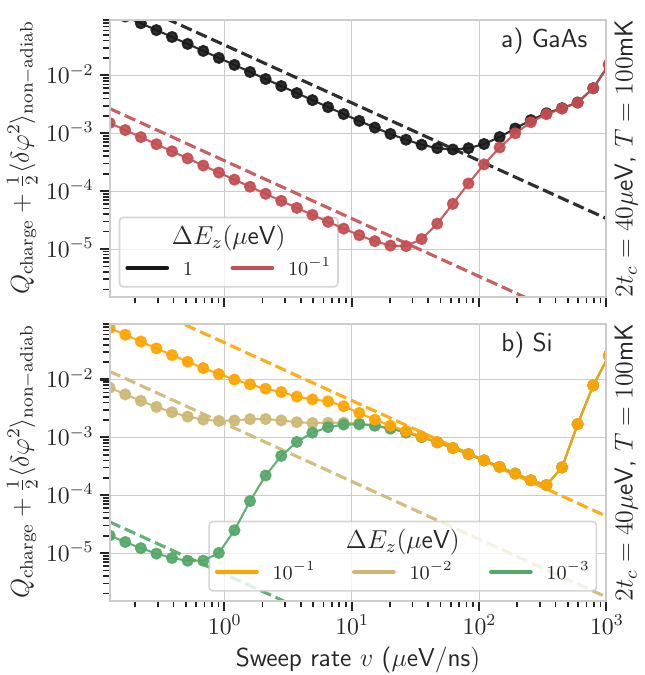}
    \caption{The sum of charge transfer error $Q_\text{charge}$ and dephasing during inelastic tunneling due to random time spent in the excited state $\tfrac{1}{2}\langle \delta \varphi^2 \rangle_\text{non-adiab}$, as a function of sweep rate $v$ for GaAs (a) and SiGe (b) devices. We show results for the tunnel coupling of $2t_c=40\mu$eV and for different values of Zeeman splittings difference $\Delta E_Z = 10^{-1}\,\mu$eV, $1\,\mu$eV for GaAs and $\Delta E_Z = 10^{-3}\,\mu$eV, $10^{-2}\,\mu$eV and $10^{-1}\,\mu$eV in Si (colors). Dots corresponds to solutions of Adiabatic Master Equation without averaging over slow noise, while dashed lines represent the approximate formula from Eq.~\eqref{eq:phi_transitions}. While at $v\sim 10^2$ $\mu$eV/ns the error is dominated by charge transfer error analyzed in Sec.~\ref{sec:charge_transfer}, at lower $v$ dephasing during inelastic tunneling dominates, as the total error becomes dependent on  the value of $\Delta E_Z$. For the case of Si and $\Delta E_Z = 10^{-1}$ (yellow  lines), the analytical formula overestimates the error since $\Delta E_Z/\Gamma_- 
    \geq 1$. We use relaxation rates from Appendix~\ref{app:relax_params}.}
    \label{fig:fast_noise}
\end{figure}

In Fig.~\eqref{fig:fast_noise}, we use dots to compare the results of AME (Eq.\eqref{eq:ame}) with approximate expression (Eq.~\eqref{eq:Qnonadiab_scaling} for GaAs (a) and Si (b) at different $\Delta E_Z$ values. The simple model assumes constant relaxation rate and spin-splitting. It is likely to overestimate the error calculated using the AME, where $\Delta E_Z$ is effectively lower around avoided crossing, i.e. when $\epsilon \ll 2t_c$. In GaAs, negligible dephasing at $\Delta E_Z \! \leqslant \! 0.1$ $\mu$eV is attributed to fast relaxation with strong electron-phonon coupling. Si exhibits lower relaxation rates, leading to longer times in the excited state. For $\Delta E_Z \! = \! 0.1\,\mu$eV (orange), the pure dephasing error combined with charge transfer error ($Q_\text{charge})$ has a single minimum, indicating significant phase loss during charge relaxation. Fig.~\ref{fig:fast_noise} reveals a region where dephasing during relaxation is ineffective, and the error is dominated by $Q_\text{charge}$. This is no longer true at slower sweeps, where the dashed lines align quantitatively with AME at $v\!<\! 50$ $\mu$eV/ns for GaAs and $v\!< \! 10$ $\mu$eV/ns for Si.

\subsubsection{Dephasing due to low-frequency (quasistatic) noise}
Second mechanism leading to spin dephasing, takes place even if the evolution is fully adiabatic, i.e. the electron stays in the ground orbital level during dot-to-dot transfer. The dephasing is then caused by low-frequency noise in the Hamiltonian parameters, which includes charge noise of a typical in a form of 1/f noise contributing to small fluctuations of detuning $\delta \epsilon(t)$ and tunnel coupling  $\delta t_c(t)$, as well as the fluctuating magnetic field due to presence of nuclear spins or charge noise induced modifications of g-factor, which we model by a dot-dependent spin-splitting noise: $\delta E_{Z,L}, \delta E_{Z,R}$. Together, the above fluctuations modify the energy splitting between the lowest energy eigenstates $\omega_\text{spin}(t) = E_{\uparrow-}(t)-E_{\downarrow-}(t)$, that defines the spin qubit energy splitting
\begin{equation}
\label{eq:spin_splitting}
    \omega_\text{spin}(t) =  E_Z + \tfrac{1}{2}[\delta E_{Z,L} + \delta E_{Z,R} -\Delta \Omega(t)],
\end{equation}
where $\Delta \Omega(t) =\Omega_{\uparrow}(t)-\Omega_{\downarrow}(t)$ is a difference in spin-depedent orbital splitting, written explicitly as
\begin{align}
\label{eq:orbital_gap_noise}
\Omega_s(t)  = \Big\{ \big[\epsilon(t)+\delta\epsilon+\tfrac{\sigma_s}{2}(\Delta E_Z+&\delta E_{Z,L}-\delta E_{Z,R}) \big]^2 \nonumber\\
& + [2t_c+2\delta t_c]^2 \Big\}^{1/2} 
\end{align}
where $\sigma_s = \pm 1$ for $s=\uparrow,\downarrow$ respectively.

The random fluctuations of spin splitting, contribute to a random phase $\delta \varphi$, since
\begin{equation}
\label{eq:varphi}
\varphi = \int_{0}^{t_f} \omega_{\text{spin}}(t) \,\text{d}t = \varphi_0 + \delta \varphi,
\end{equation}
which after averaging over realisations of all four quasistatic processes leads to dephasing $Q_\varphi \approx  \tfrac{1}{2}\langle \delta \varphi^2 \rangle$.

\begin{figure*}[t!]
    \centering
\includegraphics[width=0.99\textwidth]{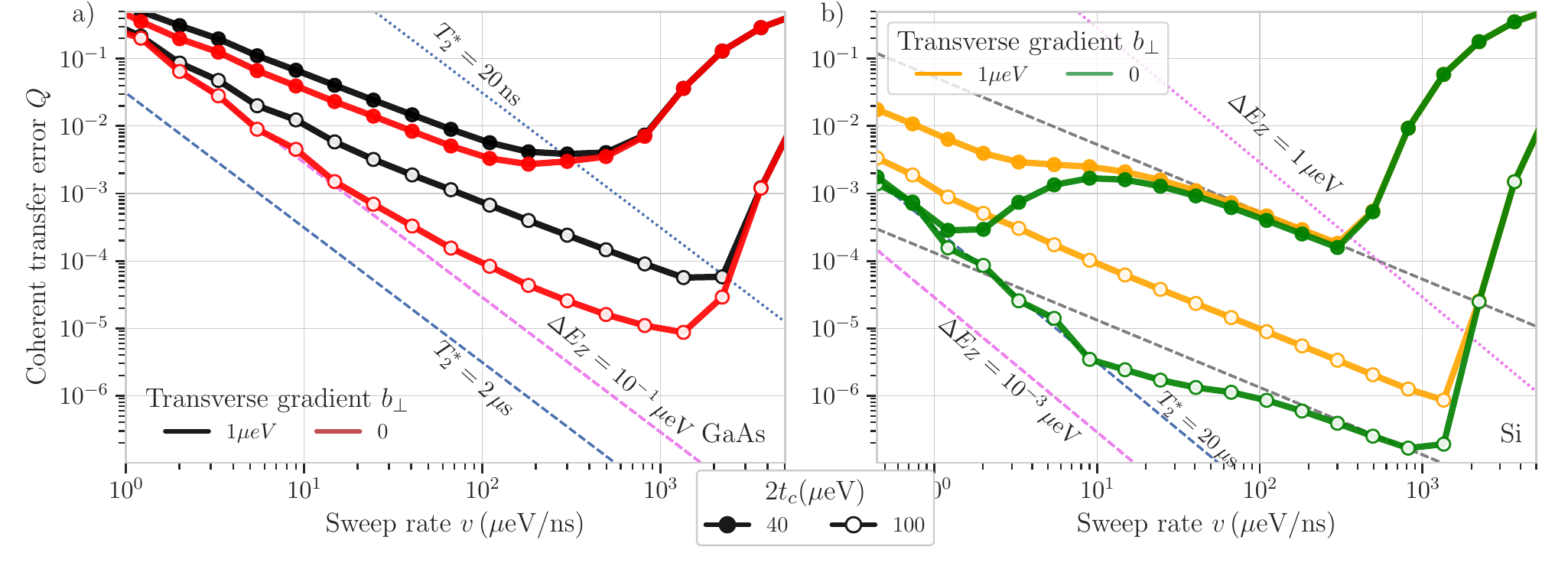}
       \caption{Total coherent transfer error $Q$ as a function of sweep rate $v$ for two values of tunnel couplings $2t_c = 40\,\mu$eV (filled dots) and $2t_c = 100\,\mu$eV (hollow dots), with results for a GaAs DQD in panel (a), and for a Si DQD in panel (b). Dots connected by lines represent solutions of the Adiabatic Master Equation averaged over quasistatic noises in spin splitting and detuning, while the dashed and dotted lines are the analytical predictions due to separate mechanisms. The violet lines are the depahsing error from Eq.~(\ref{eq:deps}) due to $1/f$ noise in detuning having an amplitude of $S_1 = 0.5$ $\mu$eV/$\sqrt{\mathrm{Hz}}$, marked by the corresponding $\Delta E_Z$, and the blue lines are phase error due to spin-splitting fluctuations in two QDs marked by corresponding $T_2^*$ values, Eq.~(\ref{eq:t2star}). For Si we also plot the phase error due to charge transfer error caused by high-frequency noise, Eq.~(\ref{eq:phi_transitions}), as gray dashed lines, corresponding to two values of $2t_c$.
       In both panels the colors of solid lines correspond to distinct values of spin-orbit couplings (SOCs). For GaAs in (a) we plot the results for intrinsic SOC $|a| = 1\,\mu$eV in red, and for synthetic SOC $b=1\,\mu$eV ($a\ll 1\mu$eV) in black. In the numerical simulation of the full error we use use $T_{2}^{*}\! =\! 2$ $\mu$s and $\Delta E_Z = 10^{-1}\mu$eV. For Si (b) we have the intrinsic SOC  $|a| = 0.1\mu$eV, and we compare the AME results for synthetic SOC $b_\perp\! = \! 0$ (green) versus $b_\perp = 1\,\mu$eV (yellow),  $T_2^* = 20\,\mu$eV and $\Delta E_Z = 10^{-3}\,\mu$eV. Influence of dephasing due to $1/f$ noise in detuning for larger values of $\Delta E_Z$ on the total error can be estimated by taking a maximum of solid line and the violet dashed line corresponding to the relevant $\Delta E_Z$. The remaining parameters can be found in Table~\ref{tab:params}.} 
    \label{fig:gaas_si}
\end{figure*} 

Calculation of $\delta \varphi$ to the lowest order in fluctuating quantitites $\delta \epsilon$, $\delta E_{Z,L(R)}$, and $\delta t_c$ is given in Appendix \ref{app:low_freq}. Let us now discuss the results for $\langle \delta \varphi^2 \rangle$ in two limits, in which it is has a simple and physically transparent form. For sufficiently large $\Delta E_Z$ the most relevant mechanism leading to non-zero $\langle \delta \varphi^2 \rangle$ are the fluctuations of dots energy detuning $\delta \epsilon$, which are directly related to the electrical noise and can be seen as a uncertainty in the exact position of the avoided crossing. In such a case similarity to the previous section, a time period spent in each dot is random, and hence $\langle \delta t^2\rangle$ becomes non-zero (see Eq.~\ref{eq:def_dt}). The value of $\delta \epsilon$ shifts initial and final values of the detuning $\pm v t_f + \delta \epsilon$, which directly translates into $\delta t \approx  \delta \epsilon/v$, and hence:
\begin{equation}
\label{eq:deps}
    \langle \delta \varphi^2\rangle = \Delta E_Z^2 \frac{\langle \delta \epsilon^2\rangle}{v^2}.
\end{equation}
Note that due to symmetry reasons contribution from slow, quasistatic fluctuations in tunnel coupling $\delta t_c$ is vanishing in the leading order, and hence does not contribute significantly to $\delta \varphi$. For more quantitative derivation of Eq.~\eqref{eq:deps} and effects of tunnel coupling noise see Appendix~\ref{app:low_freq}. We highlight that for typical amplitude of those fluctuations $\delta t_c \ll \delta \epsilon$, and in the limit of $v \ll t_c^2$ significant modification of the coherent charge excitations $Q_\text{LZ}$ are not seen in the numerical results (see \cite{Malla_PRB17} for this effect).  

In the limit of sufficiently low $\Delta E_Z$ the dephasing is dominated by standard inhomogenous broadening, caused by the fluctuations of spin-splitting. Assuming fluctuations in the dots are independent $\langle \delta E_{Z,L} \delta E_{Z,R}\rangle \!=\! 0$ and $\langle E_{Z,L}^2 \rangle= \langle \delta E_{Z,R}^2 \rangle = \sqrt{2}/T_2^*$, the contribution to a random phase for a single transition reads:
\begin{equation}
\label{eq:t2star}
\langle \delta \varphi^2 \rangle = \bigg( \frac{\Delta \epsilon}{v T_2^*}\bigg)^2,
\end{equation}
which measures the ratio of transfer time $\Delta \epsilon/v$ and typically measured dephasing time $T_2^*$. The above result holds for the nuclar spin dominated $T_2^*$, for which independence of $\delta E_{Z,L}$ and $\delta E_{Z,R}$ is expected. In the alternative scenario, when $T_2^*$ is dominated by the charge noise, e.g. g-factor fluctuations, the $\langle \delta \varphi^2 \rangle$ is expected to be slightly smaller (larger) for typically weakly correlated (anti-correlated) spin-splitting noise \cite{Yoneda_NP23, Rojas_PRAPL23, Boter_PRB20}. 

For numerical test of approximate analytical formulas, Eqs.~\eqref{eq:deps} and \eqref{eq:t2star}, see Appendix~
\ref{app:low_freq}.

\section{Case study: GaAs and Si DQD devices}
\label{sec:case_study}

We conclude this paper with the case study of the coherent double-dot transition in isotopically purified Si ($T_2^* \!= \! 20$ $\mu$s) and compare it against GaAs with continuous estimation of the Overhauser field \cite{Shulman_NC14,berritta2024real} ($T_2^* \!= \!2$ $\mu$s). The remaining parameters of the devices are discussed in Section~\ref{app:parameters}. 
The main result of the paper is shown in Fig.~\ref{fig:gaas_si}, where the dots connected with solid lines mark the coherent transfer error of equal spin superposition, computed using the Adiabatic Master Equation and averaged over low-frequency fluctuations  (see Sec.~\ref{sec:model}). For the two DQD devices, we use transition rates from Appendix~\ref{app:relax_params} and compute the transfer error $Q$ for the combination of parameters $2t_c = 40,100\,\mu$eV (filled and hollow dots respectively), and a selection of relevant parameters: single-dot spin coherence time $T_2^*=2\,\mu$s (GaAs with field narrowing), $T_2^* = 20\,\mu$s (Si), spin splitting gradient $\Delta E_Z$ between $10^{-3}$ and $1$ $\mu$eV and various spin-orbit couplings. For GaAs, we compare the relatively strong intrinsic $|a|=1\,\mu$eV (red) against spin-orbit coupling originating from the magnetic field gradient close to the manipulation region $b_\perp = 1\mu$eV (black). For Si, we use the same value of synthetic SOC (yellow) but use a  smaller intrinsic $|a|=0.1\,\mu$eV (green). We add lines corresponding to analytical predictions, dephasing noise related to spin splitting (blue, dashed), dephasing due to detuning fluctuations (violet dashed or dotted), and charge excitation using single excitation approximation limit (gray dashed). For comparison, we also add the line corresponding to $T_2^*=20$ ns without field narrowing in GaAs (dotted blue line in a).

As expected, in the limit of large sweep rates $v$, in both devices transfer error $Q$ is limited by charge excitation due to the Landau-Zener mechanism. Starting from the largest considered sweep rates, decreasing $v$ is initially improving the transfer until another mechanism becomes dominant. Those mechanisms, contrary to LZ, typically worsen the transfer as the transfer time increases, i.e. as $v$ decrease. This gives rise to an interesting trade-off mechanism we discuss below.

In GaAs, the mechanism that becomes dominant following the reduction of $v$ is spin relaxation induced by spin-orbit mixing near avoided crossings. Despite using the same value of intrinsic and synthetic spin-orbit coupling (when the latter is finite), a difference between the influence of the two couplings is visible, in particular at large $2t_c = 100\,\mu$eV, for which the intrinsic one suffers from Van Vleck cancellation since $2t_c \gg E_Z = 30\mu$eV (Red, hollow dots in the figure). The best achievable in this scenario $Q = 10^{-5}$ takes place at $v = 10^{3}\,\mu$eV/ns and increases by the order of magnitude if the SOC is of the intrinsic character. For smaller $2t_c = 40\,\mu$eV one can achieve $Q = 5\times 10^{-3}$ around $v = 10^{3}\,\mu$eV, with no visible difference between the two SOCs. At slower sweeps the $Q$ is further increasing, until the error becomes dominated by the charge noise (dashed violet line), which changes the scaling to $Q\propto 1/v^2$. 

Alternatively, at the slowest sweeps the error can be dominated by the spin splitting fluctuations in the dots. The position at which it happens depends on $T_2^*$ (see the dashed blue line for $T_2^* = 2\,\mu$s). It means that for GaAs without field monitoring ($T_2^* = 20$ns) this mechanism becomes the dominant one (see dotted violet line), and effectively creates a single minimum of $Q(v)$, with values of $Q=10^{-4}$ at $2t_c = 100\,\mu$eV , and $Q=10^{-2}$ at $2t_c = 40\,\mu$eV.

In Si, a similar  trade-off between Landau-Zener excitations, and a mechanism of dephasing due to low-frequency noise, is expected at $\Delta E_Z = 1\mu$eV (violet dotted line in b). It means that close to large longitudinal gradients (for DQD in the manipulation region) the error might not be smaller than $Q = 10^{-5}$ at $2t_c = 100\,\mu$eV and $Q = 10^{-4}$ at $2t_c = 40\,\mu$eV. Away from such region, where order of magnitude-wise $\Delta E_Z\! \sim \! 10^{-3}$, the shape of $Q$ is more complex. 

The spin relaxation mechanism in Si becomes relevant only in regions with a large transverse gradient (depicted by yellow lines with $b_\perp = 1,\mu$eV). In its presence, especially at larger $2t_c = 100,\mu$eV, the shape of the $Q$ curve (the yellow line with hollow dots) resembles the GaAs case, exhibiting a transition from Landau-Zener (L-Z) dominance at large $v$ to spin relaxation dominance at small $v$. For all the other parameter sets, a transition occurs to regime dominated by incoherent charge transfer error that is well described in the  SEAL approximation, marked by gray dashed lines. The interplay between these two charge transfer errors results in a minimum, reaching as low as $Q=10^{-7}$ at $v = 10^{3} \, \mu$eV/ns with a larger tunnel coupling of $2t_c = 100\, \mu$eV, and $Q=10^{-4}$ at $v=200\,\mu$eV/ns with smaller $2t_c=40\,\mu$eV.

Finally, as $v$ is further decreased for the Si-based DQD, $Q(v)$ curve can turn downwards again. This occurs particularly in the absence of strong synthetic SOC (green lines), and manifests at $v$ small enough to allow for relaxation from the excited to the target state, i.e.~the HEAL process from Sec.~\ref{sec:charge_transfer}. This process can temporarily decrease the error for smaller $v$, until the dephasing due to low-frequency noise in $\epsilon$ or spin splitting become dominant at lowest $v$ due to $Q = 1/v^2$ scaling. Depending on the exact values of $T_2^*$ (blue line), charge noise amplitude and $\Delta E_Z$ (violet dashed and dotted line), it can give rise to a second minimum of $Q$ at low $v$, which happened in case of $2t_c =40\,\mu$eV (green line with hollow dots). However, the error at this local minimum, $Q \! \approx\! 5\times\,10^{-4}$, is slightly larger than the one at local minimum at larger $v$, and 
and is hardly visible at larger $2t_c =100\,\mu$eV.

\section{Implications for transfer of a qubit through a chain of dots}
\label{sec:discussion}
\begin{figure}
    \includegraphics[width = \columnwidth]{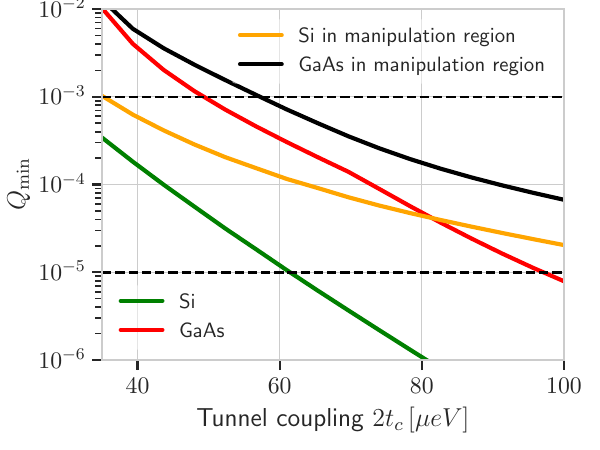}
    \caption{Estimated lowest coherent transfer error $Q_\text{min}$ between neighouring dots, as a function of tunnel coupling, for the case of dots away from manipulation region in Si (green) and GaAs (red) quantum dots, compared against the same dots in the manipulation region depicted using yellow and black color, respectively. In the first case we use parameters from Table~\ref{tab:params}, and for manipulation region we set $b_\perp = \Delta E_Z = 1\,\mu$eV. The dashed horizontal lines denotes the threshold for the single weak link, $Q \approx 10^{-3}$, and for the uniform array of $N=100$ dots, $Q = 10^{-5}$.  }
     \label{fig:funtunc}
\end{figure}

We summarize the findings in Fig.~\ref{fig:funtunc}, where we plot the predicted minimum of the coherent transfer error $Q$ as a function of tunnel coupling. For both devices, we maintain the device parameters as in Fig.~\ref{fig:gaas_si}, and we compare two cases: one where the transition occurs close to the manipulation region with $\Delta E_Z = b_\perp = 1\,\mu eV$, and another where the transition is far from the manipulation region, in case of which we set the intentional gradient to zero. In this Section, the results shown in this Figure will be used to analyze the prospects for coherent spin transfer through a long ($\sim 1-10\,\mu$m) array of tunnel-coupled quantum dots. 

We begin by discussing the mechanisms of charge transfer error and spin relaxation, which dominate the lowest achievable error $Q$, for the whole range of considered $2t_c>E_Z= 30\mu$eV. For these processes, the total error accumulated during transfer is the sum of the contributions from each dot-to-dot transfer, i.e.,
\[
    Q_{N} \approx \sum_{n=1}^{N} Q_\text{charge}^{(n)} + Q_\text{relax}^{(n)},
\]
where $Q^{(n)}$ represents the error accumulated during the transition between the $(n-1)$th and $n$th dot. A single charge-transfer error implies that the electron is left behind in the initial dot when it should have transferred to the next one. However, in the simplest version of a scalable shuttling scheme —  in which a  few global voltages are used to simultaneously trigger the appropriate interdot detuning swings in a long chain of quantum dots (QDs) \cite{Langrock_PRXQ23}—an electron left behind will begin traveling in the opposite direction in subsequent driving cycles. In a more complex scheme, it would resume moving in the correct direction along the QD chain. Nevertheless, even in this latter scenario, the delay caused by staying in a random dot (with a random deviation of Zeeman splitting from the average value) for one period of driving leads to significant dephasing of the delayed qubit. For simplicity, we assume here that each charge transfer failure results in an irreversible loss of coherence for the shuttled qubit.

For a QD chain with uniform parameters, for which each pair of adjacent QDs has $2t_c^{(n)} \approx 2t_c^{(0)}$, the overall transfer error is estimated as $Q_{N} = N Q_1$, where $Q_1$ is the error through typical transition away from manipulation region. With these assumptions, the coherent transfer of $Q_N < 10^{-3}$ through $N = 100$ dots corresponding to approximately shuttling distance of $L = 10$ $\mu$m \cite{Boter21}, would require $2t_c >60$ $
\mu$eV in Si and $2t_c > 100$ $\mu$eV in GaAs, as illustrated by the lower dashed line in Fig.~\ref{fig:funtunc}. From our results it is clear that tunnel coupling below that value, would not allow for a transfer across  $L \!= \!10$ $\mu$m even in a perfectly uniform QD chain. 

Due to nature of the errors, the corresponding optimal sweep rate for the coherent transfer aligns with the one that minimises the probability of excitation to the higher energy state, making the tunnel coupling requirements across the array consistent with those outlined in \cite{Krzywda_PRB21}. In that study, in absence of spin degree of freedom, it was suggested that slower sweep rates could enhance charge transfer. However, here we have shown that this strategy is unlikely to be beneficial for coherent transfer, as lowering $v$ below the value optimal for successful charge transfer triggers significant spin dephasing and relaxation.

On the other hand, when tunnel couplings between neighboring dots in a long chain 
exhibit significant spread around their mean value (assumed now to be larger than the above-discussed limit of $2t_ c \!= \! 60$ $\mu$eV) the coherent transfer error can be limited by a few weak links (dots coupled by smallest $t_c$, or dots in manipulation region coupled with not-large-enough $t_c$), or in an extreme case, by a single transition with largest error $Q$, i.e. $Q_N \approx \text{max}(Q)$. Based on the results from Fig.~\ref{fig:funtunc} this puts constrains on the weakest tunnel coupling along the chain to $2t_c \!>\! 50\,\mu$eV in GaAs away from manipulation region (red line), which is needed for $Q \!>\! 10^{-3}$. In  manipulation regions, where gradients and the $g$-factor difference are likely to be significantly larger, the minimal tunnel coupling is higher and equal to $2t_c \!>\! 60\,\mu$eV in GaAs and $2t_c\! > \! 35\,\mu$eV in Si, as illustrated in Fig.~\ref{fig:funtunc} by the intersection of the upper dashed line with the red and yellow lines respectively. 


While for assumed magnetic field of $E_Z = 30\mu$eV, the optimal $Q$ was limited by the charge and relaxation errors, for smaller magnetic field a second minimum created at much lower sweep rate can give a lower $Q$ (see a case of two minima in Fig.~\ref{fig:gaas_si} b). If the left minimum is lower, the value of lowest $Q$ becomes dominated by the dephasing caused by spatiotemporally correlated low-frequency noise, for which additivity might no longer hold. In particular, assuming dephasing is caused by the coherent sum of random phases:
\begin{equation}
    \delta \varphi_N = \sum_{n=1}^N \delta \varphi^{(n)},
\end{equation}
contribution to an error after averaging depends on the correlations $\langle \delta \varphi^2 \rangle = \sum_{nm} \langle \delta \varphi^{(n)}\delta \varphi^{(m)}\rangle$. For a linear chain, each dot is visited only once, which is expected to introduce only a small modification to linear scaling due to only weak correlation of the noise in nearest neighbours $ \langle \delta \varphi^{(n)}\delta \varphi^{(n+1)}\rangle \ll  \langle [\delta \varphi^{(n)}]^2\rangle $. Assuming the uniform chain of dots $\langle [\delta \varphi^{(n)}]^2\rangle \approx  Q_{\varphi,1}$ we can estimate
\begin{equation}
    \langle \delta \varphi_N^2 \rangle \approx \xi  N Q_{\varphi,1}
\end{equation}
where numerical factor $\xi \gtrsim 1$ encapsulates spatial correlation of charge noise between nearest and next-nearest neighbours \cite{Rojas_PRAPL23}. The above is the manifestation of motional narrowing process, which in comparison to stationary system, decreases coherent error by the virtue of interacting with many uncorrelated environments \cite{abragam1961principles}. Such motional narrowing leading to decreased dephasing of moving qubit compared to the stationary one was observed in recent shuttling experiment with a moving QD in Si/SiGe \cite{Struck_NC24}.

In light of recent experimental results in Ge hole systems \cite{VanRiggelen_NC24}, we emphasize that correlations of low-frequency noise are expected to be more relevant for sequential transitions between a few quantum dots. For instance in Ref.~\cite{VanRiggelen_NC24} the tunnel coupling \( 2t_c \approx 100 \, \mu \text{eV} \) and optimised sweep rate should prevent significant charge excitations. Instead, we expect the error to stem from relaxation or, more likely,  dephasing caused by the large \(\Delta E_Z\) specific to holes in Ge \cite{Hendrickx_Nature20, Hendrickx_NC20, Jirovec_PRL22}. However, in a setup where the shuttling occurs back and forth between two or three dots, the same dot is visited multiple times. This means that, even without spatial correlations in the noise, slow fluctuations \(\delta \varphi\) can accumulate coherently, potentially resulting in \(\langle \delta \varphi^2 \rangle > N^\alpha\) scaling, with \(1 < \alpha < 2\). Moreover, repeating the shuttling process could create a noise filter that is commensurate with the shuttling frequency. Thus, predicting the exact relationship between the spectral density of the noise, the electron path, and the resulting dephasing requires a separate, detailed analysis, which lies outside the scope of this paper.


\section{Summary}
In conclusion, we have analyzed the most relevant mechanisms leading to errors in coherent electron spin qubit transfer between tunnel-coupled quantum dots in Si and GaAs. Although charge transfer error in the presence of realistic environmental noise can be made arbitrarily small by using small enough sweep rates $v$ \cite{Krzywda_PRB21}, the same does not hold for the coherent transfer error, $Q$, to which spin relaxation and dephasing also contribute. This happens because all the discussed here mechanisms of spin dephasing and relaxation become more efficient with decreasing $v$. Channels for spin dephasing associated with interdot qubit transfer are activated in presence of nonzero difference in Zeeman splitting, $\Delta E_Z$, in the two dots (due to either a $g$-factor difference, or presence of magnetic field gradient) and randomness in fractions of time spent in each of the dots caused by interaction of electron charge with high- and low-frequency electric field noise. The high-frequency noise can excite the electron out of its instantaneous ground state when the energy gap between this state and the excited one is minimal - i.e.~at the tunnel-coupling induced anticrossing. Even if the electron eventually relaxes due to phonon emission to the correct final state in the target dot (so that the charge transfer error is much lower that the probability of the excitation at the anticrossing), the stochastic character of this relaxation leads to a random contribution to spin phase, and thus to spin qubit dephasing. 
Spin relaxation is also enhanced compared to the relaxation in one of the dots when the electron is delocalized between the dots, and the mixing of ground and excited state by spin-orbit interaction (either intrinsic, or extrinsic, due to transverse component of the mangetic fild gradient) is the strongest due to the energy gap between these states being the lowest. 
Finally, the low-frequency noise in detuning gives a random contribution to phase of the qubit even when charge dynamics is perfectly adiabatic, and averaging over this noise give phase error that scales as $\Delta E_Z^2/v^2$. Clearly, large $\Delta E_Z$, and small tunnel coupling $t_c$ (which, apart from enhancing excitation probability and spin relaxation at the anticrossing, necessitates smaller $v$ to avoid nonadiabaticity of charge dynamics) are detrimental for coherent spin transfer. 

In both GaAs and Si we have found a nonmonotonic dependence of $Q$ on $v$, with the lowest $Q$ at an optimal $v$. For GaAs in which the otherwise dominant dephasing due to hyperfine interaction with nuclei was weakened by monitoring of slow fluctuations of nuclear Overhauser field \cite{Klauser_PRB06,Bluhm_PRL10,Shulman_NC14,berritta2024real}, the most important shuttling-induced decoherence process is spin relaxation near the anticrossing. For Si, $Q$ sometimes exhibits two local minima, however, for most values of parameters considered here the global minimum of $Q$ occurs when processes of coherent and incoherent charge transfer error become comparably effective, i.e.~the minimal coherent transfer error coincides with the minimal charge transfer error from \cite{Krzywda_PRB21}. A notable exception is the case of large $\Delta E_Z$, for which it is the competition between Landau-Zener process, and dephasing due to low-frequency noise in detuning that determines the optimal $v$ and minimal $Q$.


Applying our results to the case of shuttling for $\sim \! 10$ $\mu$m distance (i.e.~through a chain of $\sim \! 100$ tunnel-coupled quantum dots), we have found that in  silicon, coherent transfer with error $\leq \! 10^{-3}$ will require \( 2t_c > 60 \, \mu \text{eV} \) for most dot pairs. When ``weak links'' with tunnel couplings twice smaller ($2t_c \! \approx \! 30$ $\mu$eV) are present, qubit transport through them will dominate the total error and make it larger than $10^{-3}$. Shuttling through the ``manipulation zones'' near micromagnets is most error-prone, and weak links in these zones will be most detrimental for coherent shuttling. All these observations requirements for a hundred-dot array make the bucket brigade method of long-distance qubit transfer challenging for future designs, unless a method to obtain large and uniform tunnel coupling ($2t_c$ equal to at least 60 $\mu$eV), without having to tune up each tunnel barrier separately, is found.

\section*{Code availability}
The code used to generate numerical results in this study has been made available at \cite{github}.
\appendix

\acknowledgements
This project was supported by the European
Union's Horizon 2020 research and innovation programme
under grant agreement 101174557 (project QLSI$^{2}$).

\section{Lindblad form of the Adiabatic Master Equation}
\label{app:AME}

In this appendix we derive Lindblad form of the Adiabatic Master Equation provided in Eq.~\eqref{eq:ame}.
\subsection{Instantaneous states, energies and adiabatic frame}
\label{sec:adiabatic_states}
We start by considering the spin-diagonal Hamiltonian 

\begin{equation}
        \hat H(t) = t_c\hat \tau_x + \frac{\epsilon(t)}{2}\hat \tau_z + \frac{E_Z}{2}\hat \sigma_z + \frac{\Delta E_Z}{4} \hat \tau_z \hat \sigma_z.
    \end{equation}
    where $2t_c$ is the tunnel coupling, $\epsilon(t)$ is the dots deutning, $\Delta E_Z = E_{Z,L}-E_{Z,R}$ and $E_Z = \tfrac{1}{2}( E_{Z,L}+E_{Z,R})$ are related to dot-dependent spin-splittings $E_{Z,L}, E_{Z.R}$. 
    
    At any point in time, the above Hamiltonian can be diagonalized via the rotation around the y-axis:
${\hat A(t) \!=\! \sum_{s=\uparrow,\downarrow} e^{-i\theta_s(t)\hat \sigma_y/2} \ketbra{s}}$ using spin-dependent orbital angle defined via $\text{ctan}\,\theta_s(t) = [\epsilon(t) + \sigma_s\Delta E_Z/2]/2t_c$ where $\sigma_s = \pm 1$ are for spin-up and spin-down respectively. After diagonalization we have:
    \begin{equation}
        {\tilde H}_0(t) = \hat A^\dagger(t) \hat H(t) \hat A(t) =  \frac{1}{2}\sum_{s=\uparrow,\downarrow} \bigg(\Omega_s[t] \adiab{\tau}_z + \sigma_s E_Z\bigg) \ketbra{s},
    \end{equation}
    where $\Omega_s[t] = \sqrt{(2t_c)^2 + (\epsilon[t]+\sigma_s \Delta E_Z/2)^2}$ is the spin-dependent orbital gap, while $\adiab{\tau}_z$ is the Pauli matrix in the \textit{adiabatic frame}, in which only coefficients of the system state vary in time, and the corresponding state vectors are time-independent. In such an \textit{adiabatic frame} we define the state $|\tilde \psi(t)\rangle$ via the transformation:
\begin{align}
\label{eq:adiab}
\ket{\psi(t)} &= \sum_s a_{s+}(t)\ket{\psi_{s+}(t)} + a_{s-}(t) \ket{\psi_{s-}(t)} \nonumber \\&=  \hat A(t) \bigg[\sum_s \tilde a_{s+}(t)\ket{\psi_{s+}(t_0)} + \tilde a_{s-}(t) \ket{\psi_{s-}(t_0)}\bigg] \nonumber \\&= \hat A(t) |\tilde \psi(t)\rangle.
\end{align}
When $|\tilde \psi(t)\rangle = \hat A^\dagger(t) \ket{\psi(t)}$ is substituted to Schroedinger equation, we have
$  i \partial_t\big(\hat A(t)|\tilde \psi(t)\rangle\big) = \hat H_0(t) A(t)|\tilde \psi(t)\rangle$, which multiplied from the left by $\hat A^\dagger(t)$ produces the Schroedinger equation in the \textit{adiabatic frame}:
\begin{equation}
    i\partial_t |\tilde \psi(t)\rangle = \left(\adiab H_0(t) -i \hat A^\dagger(t) \partial_t \hat A(t) \right) |\tilde \psi(t)\rangle = \adiab{H}|\tilde \psi(t)\rangle.
\end{equation}
In the above equation we obtained the coupling between adiabatic levels,
\begin{equation}
\hat A^\dagger(t) \partial_t \hat A(t) = \frac{1}{2}\sum_{s=\uparrow,\downarrow} \partial_t \theta_s(t) \adiab\tau_y \ketbra{s},
\end{equation}
which for the linear drive $\epsilon(t) = v t$ reads $\partial_t \theta_s = 2v t_c / \Omega_s^2(t)$. 

\subsection{Adiabatic Bloch-Redfield equation}
We now add environment in thermal equilibrium with environmental Hamiltonian $\hat H_e$, i.e. $\hat \varrho_e = e^{-\beta\hat H_e}/\mathcal{Z},$ where $\mathcal Z = \text{Tr}\{e^{-i\hat H_e}\}$ and the inverse thermal energy is $\beta = 1/k_B T$. The environment couples only to the charge states of the electron, i.e. 
\begin{equation}
\hat V_e = \tfrac{1}{2}\left(\hat V_{x} \hat \tau_x +  \hat V_z \hat \tau_z\right)
\end{equation} 
where $\hat V_i$ are the real operators acting on environmental degrees of freedom only. 

The von Neumann equation for the density matrix of the qubit-environment system in the adiabatic frame reads
\begin{equation}
    \frac{\partial}{\partial t}  {\tilde\varrho}_{qe}(t) = -i \comm{{\tilde H}(t)+  {\tilde V}_e(t)+ \hat H_e}{ {\tilde \varrho}_{qe}(t)},
\end{equation}
where we have used the operators in the \textit{adaibatic frame} i.e. ${\tilde O}(t) = \hat A^\dagger(t) \hat O \hat A(t)$. We now follow standard treatment, i.e. perform Born-Markov approximation that leads to the Bloch-Redfield equation in the interaction picture w.r.t. $\adiab H(t) + \hat H_e$, i.e.
\begin{align}
\frac{\partial}{\partial t}  {\tilde\varrho}_{q,I}(t) = - \Tr_{e}&\left\{\comm{{\tilde V}_{e,I}(t)}{\int_{0}^{\infty} \text{d}s\, {\tilde V}_{e,I}(t-s){\hat \varrho}_{q,I}'(t) \hat \varrho_e}\right\} \nonumber \\ &\qquad+ (h.c.),
\end{align}
in which we have defined $${\tilde O_I(t) = e^{iH_e t} \hat U_q^\dagger(t,t_0) \tilde O U_q(t,t_0) e^{-iH_e t}},$$ with $
{\tilde U_q}(t,t_0) =  \mathcal{T}\exp{-i\int_{t_0}^t  {\tilde H}(t') \text{dt'}}$.

The Adiabatic Master Equation is now obtained by unwinding the interaction picture only with respect to the system, i.e. multiplying from the left by $\hat U_q(t,t_0)$ and from the right by $\hat U_q^\dagger(t,t_0)$, which after simplifications gives:
\begin{align}
    \frac{\partial}{\partial t}{\tilde \varrho}_q(t) = &-i\comm{ {\tilde H}(t)}{{\tilde \varrho}_{q}(t)}-\Tr_e\bigg\{ \bigg[{ {\tilde V}_{e,I_e}}(t),\int_{0}^{\infty} \text{d}s \nonumber \\ &\Big({\tilde U}(t,t-s){{ \tilde V}_{e,I_e}}(t-s) { \tilde U}^\dagger(t,t-s)\Big) \hat\varrho_q'(t)\bigg] \nonumber \\&+ (h.c.) \bigg\},
\end{align}
where  ${\tilde {V}}_{e,I_e}(t) = e^{i \hat H_e t} \hat A^\dagger(t)\hat {V}_e(t) \hat A e^{-i \hat  H_e t}$ is the operator in the adiabatic frame and in the interaction picture with respect to environmental Hamiltonian only. 

We now perform the central approximation of adiabatic master equation, which is related to a free-evolution operator inside the dissipative part, which is approximated as: 
\begin{equation}
\tilde {U}(t,t-s) =  \mathcal {T} \exp(-i \int_{t-s}^{t}  \adiab{H}(t') ) \approx \exp(-i  {\tilde H}_0(t)\, s ).
\end{equation} 
That is equivalent of saying that within the short correlation time of the environment $s<\tau_c$, both coherent (adiabatic) coupling between the adiabatic levels as well as a rotation of the adiabatic frame can be neglected. The latter also means that we can write:
\begin{align}
    \hat{U}(t,t-s)& {\tilde V}_{e,I_e}(t-s)\hat{ U}^\dagger(t,t-s)  \\& \approx  e^{-i{\tilde H}(t)\, s }   \hat A^\dagger(t) \hat V_{e,I_e}(t-s)\hat A(t) e^{i {\tilde H}(t)\, s }  \nonumber
\end{align}
where $\hat V_{e,I_e}(t) = e^{i\hat H_e t}\hat V_e e^{-i\hat H_e t}$. It is now convenient to introduce the resolution of the unity ${\sum_n \ketbra{n(t_0)} \approx \sum_n \tilde\Pi_n}$ in terms of time-independent states in the adiabatic picture, use $\hat A(t)\ket{n(t_0)} = \ket{n(t)}$, $\tilde H_0(t)\ket{n(t_0)} = E_n \ket{n(t_0)}$ and arrive at:
\begin{align}
\sum_{nm} & e^{-i{\tilde H}(t)\, s }  \tilde \Pi_n  \hat A^\dagger(t) \hat V_{e,I_e}(t-s)  \hat A(t) \tilde \Pi_m e^{i {\tilde H}(t)\, s }  \\&= \sum_{nm} \bra{n(t)}V_{e,I_e}(t-s)\ket{m(t)} e^{-iE_{nm}s} \tilde \Pi_{nm} \nonumber
\end{align}
where $E_{nm}=E_n(t)-E_m(t)$ is the difference in the energies of the adiabatic states $\ket{n(t)}, \ket{m(t)}$ and $\tilde \Pi_{nm} = \ketbra{n(t_0)}{m(t_0)}$.  

Together the adiabatic Master equation for the density matrix of the electron state now reads:
\begin{align}
\label{eq:nmlk}
    &\frac{\partial}{\partial t}{\tilde \varrho}_q(t) = -i\comm{ {\tilde H}(t)}{ {\tilde \varrho}_{q}(t)}  \\& - \sum_{nmlk}\comm { {\tilde \Pi}_{nm}} { {\tilde \Pi}_{kl} {\tilde \varrho}_{q}(t)\int_{0}^{\infty} \text{d}s \, e^{-iE_{kl}[t]s} A_{nmkl}(s) } + (h.c.) \nonumber,
\end{align}
in which we have defined 
\begin{align}
    A_{nmkl}(s) &=  \Tr{\hat V_{e,I_e}^{nm}(t) \hat V_{e,I_e}^{kl}(t-s)\hat \varrho_e} \nonumber \\&= 
    \Tr{\hat V_{e,I_e}^{nm}(s) \hat V_{e,I_e}^{kl}(0)\hat \varrho_e}
\end{align}
with $\hat V_{e,I_e}^{nm}(t) = \bra{n(t)}\hat V_{e,I_e}(t)\ket{m(t)}$ for brevity. We now substitute $\hat V_e = \frac{1}{2}\sum_{i=x,z} \hat \tau_i \hat V_i$ a which convenient gives:
\begin{equation}
    A_{nmkl}(s) = \sum_{i=x,y}  \tau_{i,nm}(t) \tau_{i,kl}(t) C_i(s),
\end{equation}
where $\tau_{i,nm}(t) = \bra{n(t)} \hat \tau_i \ket{m(t)}$ and $C_i(s) = \Tr\{\hat V_i(s) \hat V_i(0) \hat \varrho_e\}$ is the correlation function, in which we have assumed that two environmental operators are statistically independent, i.e. ${\langle \hat V_z \hat V_x\rangle = \Tr_e\{\hat V_z \hat V_x \hat \varrho_e\} = 0}$. We finally express the correlation function in terms of spectral density of the noise, i.e. $C_i(r) = \int_{-\infty}^{\infty} S_i(\omega) e^{-i\omega r} \frac{\text{d}\omega}{2\pi} $, which substituted to the expression in the commutator reads:
\begin{align}
    &\int_{0}^{\infty} \text{d}s \, e^{-iE_{kl}[t]s} C_{nmkl}(r) = \sum_{i=x,z} \frac{\tau_{i,nm}(t)  \tau_{n,kl}(t)}{8} S_i(E_{lk}[t]), 
\end{align}
In the above, we have used the identity $\int_0^{\infty} e^{i\omega t}\text{d}t = \pi \delta(\omega) + P(1/\omega)$ and ignored the principle value as it generates only deterministic shift in the energy. In thermal equilibrium we have $S(\omega) = S(-\omega)e^{-\beta \omega}$ where $\beta = 1/k_\text{B}T$.

\subsection{Local secular approximation}
In the dissipative part of Eq.~\eqref{eq:nmlk} there is a sum over four different adiabatic states. However some of their combination include quickly oscillating terms, which will not contribute strongly to the $\tilde \varrho_q$. To see this mechanism we go back to the interaction picture with respect to $\hat H_0$, i.e.:
\begin{align}
    \frac{\partial}{\partial t}&{\tilde \varrho}_{q,I}(t) = - i \comm{\hat U_0^\dagger(t,t_0) A^\dagger(t)\partial_t A(t)\hat U_0(t,t_0)}{\varrho_{q,I}(t)} \nonumber \\&- \frac{1}{8}\sum_{nmkl} e^{-i\int_0^{t}(E_{nm}[t']+E_{kl}[t'])\text{d}t'} \comm { {\tilde \Pi}_{nm}} {\tilde { \Pi}_{kl}\hat { \varrho}_{q,I}(t)}  \nonumber \\ & 
\times \sum_{i=x,z}\tau_{i,nm}(t)\tau_{i,kl}(t) S_i(E_{lk}[t]) + (h.c.).
\end{align}
We now attempt to neglect those highly oscillating terms. Neglecting all of the terms where $E_{nm}[t'] \neq -E_{kl}(t)$, i.e. when $n=l$ and $m=k$, we correspond to a global secular approximation, that leads to a Lindlbad equation. However such an approximation would not preserve any coherence between spin-up and spin-down states during their orbital dissipation.

Intuitively the amount of spin coherence lost during the orbital relaxation depends on the ratio between $\Delta E_Z/\Gamma_\pm$, i.e. slow relaxation would randomize time spend in two orbital eigenstates of different spin splitting. Since typicall $\Delta E_Z/\Gamma_\pm$ is small, we keep the terms where $|E_{nm}[t']-E_{kl}[t']| \leq \Delta E_Z$, i.e. we perform a local secular approximation with respect to the orbital gap \cite{winczewski2024intermediate}. As a result there are only 20 non-vanishing combinations of indices, which are summarized in Table~\ref{table:transitions}. 

\begin{table}[htb!]
\begin{tabular}{c|c|c|c|c|c|c|c|}
\cline{2-8}
\multicolumn{1}{l|}{}     & \textbf{n}     & \textbf{m}               & \textbf{k}               & \textbf{l}               & \textbf{$E_{nm}+E_{kl}$}  & \textbf{$ E_{kl}$}        & SOC needed \\ \hline
\multicolumn{1}{|c|}{c}  & $s-$           & $s+$                     & $s+$                     & $s_-$                    & 0                         & $\Omega_s$                & no         \\ \hline
\multicolumn{1}{|c|}{c}  & $s+$           & $s-$                     & $s-$                     & $s+$                     & 0                         & $-\Omega_s$               & no         \\ \hline
\multicolumn{1}{|c|}{c}  & $s-$           & $s+$                     & $\overline s+$           & $\overline s-$           & $-\sigma_s \Delta \Omega/2$ & $\Omega_{\overline{s}}$     & no         \\ \hline
\multicolumn{1}{|c|}{c}  & $s+$           & $s-$                     & $\overline s -$          & $\overline s+$           & $\sigma_s \Delta \Omega/2$  & $-\Omega_{\overline{s}}$    & no         \\ \hline
\multicolumn{1}{|c|}{s}  & $\uparrow q$   & $\downarrow q$           & $\downarrow q$           & $\uparrow q$             & 0                         & $-E_Z - q\Delta\Omega$    & yes        \\ \hline
\multicolumn{1}{|c|}{s}  & $\downarrow q$ & $\uparrow q$             & $\uparrow  q$            & $\downarrow q$           & 0                         & $E_Z + q\Delta \Omega/2$    & yes        \\ \hline
\multicolumn{1}{|c|}{s}  & $\uparrow q$   & $\downarrow  q$          & $\downarrow \overline q$ & $\uparrow \overline q$   & $q\Delta \Omega/2$          & $-E_Z  +q \Delta \Omega/2$  & yes        \\ \hline
\multicolumn{1}{|c|}{s}  & $\downarrow q$ & $\uparrow q$             & $\uparrow \overline q$   & $\downarrow \overline q$ & $-q\Delta \Omega/2$         & $E_Z - q\Delta \Omega/2$    & yes        \\ \hline
\multicolumn{1}{|c|}{cs} & $\uparrow q$   & $\downarrow \overline q$ & $\downarrow\overline q$  & $\uparrow q$             & 0                         & $-E_Z +q\overline \Omega$ & yes        \\ \hline
\multicolumn{1}{|c|}{cs} & $\downarrow q$ & $\uparrow \overline q$   & $\uparrow \overline q $  & $\downarrow q$           & 0                         & $E_Z-q \overline \Omega$  & yes        \\ \hline

\end{tabular}

\caption{Relevant transition rates, that survives local secular approximation. We use the notation in which $\sigma_s = \pm 1$ for spin-up and spin-down respectively while $q = \pm 1$ for excited and ground orbital state. In the dissipation part we neglect difference in the relaxation rates that are of the order of $\Delta \Omega < \Delta E_Z \ll \Omega_0(t)$.}
\label{table:transitions}
\end{table}

Additionally since $\Delta E_Z$ is also much smaller than any other timescale, in the dissipative part we neglect the difference between spin-diagonal orbital splittings $\Delta \Omega = \Omega_\uparrow - \Omega_\downarrow \to 0$, and set:
\begin{equation}
    \Omega_\uparrow(t) \approx \Omega_\downarrow(t) \approx \Omega_0(t) = \sqrt{\epsilon^2(t) + (2t_c)^2},
\end{equation}
which also means that $\overline \Omega = \Omega_0$. Note that this is consistent with local secular approximation, as with this assumptions all of the transition listed in Table~\ref{table:transitions} have $E_{nm} + E_{kl} = 0$. Additionally, what follows is that all of the charge transitions (c) are associated with the same energy transfer $\pm \Omega_0$, the spin transitions (s) with common energy transfer of $\pm E_Z$, while combined charge-spin transitions with $\pm (\Omega_0-E_Z)$ (cs1) and $\pm (\Omega_0+E_Z)$ (cs2). 

In the same limit of small $\Delta E_Z$, the matrix elements can be approximated as their value at $\Delta E_Z = 0$, for which the spin-up and spin-down transitions are alligned and $\theta_\uparrow = \theta_\downarrow = \theta_0$. This would mean:
\begin{equation}
    \bra{\psi_{s+}(t)}\hat \tau_i \ket{\psi_{s-}(t)} \approx \bra{+(t)} \hat \tau_i \ket{-(t)},
\end{equation}
where $\bra{+(t)} \hat \tau_z \ket{-(t)} = -\cos\theta_0$, $\bra{+(t)} \hat \tau_x \ket{-(t)} = \sin\theta_0$ and $\text{ctan}\theta_0 = \epsilon(t)/2t_c$.

As a result of the assumptions above, the final form of Adiabatic Master Equation has the Lindblad form, i.e. it can be written as:
\begin{align}
\label{eq:ame_app}
&\frac{\partial}{\partial t}  {\tilde \varrho}_q(t) = - i\comm{ {\tilde H}(t)}{{\tilde \varrho}_q(t)} + \sum_{q = \pm } \bigg[\\\nonumber &+  \bigg({\tilde L}_{q,c}(t) {\tilde \varrho}_q(t) \hat {\tilde L}_{q,c}^\dagger(t)  - \frac{1}{2}\acomm{ {\tilde L}_{q,c}^\dagger(t)  {\tilde L}_{q,c}(t)}{{\tilde \varrho}_q(t)}\bigg)\\\nonumber &+   \bigg({\tilde L}_{q,s}(t) {\tilde \varrho}_q(t) \hat {\tilde L}_{q,s}^\dagger(t)  - \frac{1}{2}\acomm{ {\tilde L}_{q,s}^\dagger(t)  {\tilde L}_{q,s}(t)}{{\tilde \varrho}_q(t)}\bigg)\\\nonumber &+ \bigg({\tilde L}_{q,cs1}(t) {\tilde \varrho}_q(t) \hat {\tilde L}_{q,cs1}^\dagger(t)  - \frac{1}{2}\acomm{ {\tilde L}_{q,cs1}^\dagger(t)  {\tilde L}_{q,cs1}(t)}{{\tilde \varrho}_q(t)}\bigg)\\\nonumber &+ \bigg({\tilde L}_{q,cs2}(t) {\tilde \varrho}_q(t) \hat {\tilde L}_{q,cs2}^\dagger(t)  - \frac{1}{2}\acomm{ {\tilde L}_{q,cs2}^\dagger(t)  {\tilde L}_{q,cs2}(t)}{{\tilde \varrho}_q(t)}\bigg)\bigg],
\end{align}
in which we defined time-dependent Lindblad operators of explicit form:
\begin{align}
{\tilde L}_{\pm,c}(t) &=    \sqrt{\Gamma_{\pm,c}(t)}\sum_{s=\uparrow,\downarrow} |\tilde \psi_{s\pm}\rangle \langle \tilde \psi_{s\mp}|  \nonumber \\
{\tilde L}_{-,s}(t) &=    \sqrt{\Gamma_{-,s}(t)}\sum_{q=\pm} |\tilde \psi_{\downarrow q}\rangle \langle \tilde \psi_{\uparrow q}|  \nonumber \\
{\tilde L}_{+,s}(t) &=    \sqrt{\Gamma_{+,s}(t)}\sum_{q=\pm} |\tilde \psi_{\uparrow q}\rangle \langle \tilde \psi_{\downarrow q}|  \nonumber \\
{\tilde L}_{-,cs1}(t) &=    \sqrt{\Gamma_{-,cs1}(t)}\, |\tilde \psi_{\uparrow -}\rangle \langle \tilde \psi_{\downarrow +}|  \nonumber \\
{\tilde L}_{+,cs1}(t) &=    \sqrt{\Gamma_{+,cs1}(t)} \,|\tilde \psi_{\downarrow +}\rangle \langle \tilde \psi_{\uparrow -}|  \nonumber \\
{\tilde L}_{-,cs2}(t) &=    \sqrt{\Gamma_{-,cs2}(t)}\, |\tilde \psi_{\downarrow -}\rangle \langle \tilde \psi_{\uparrow +}|  \nonumber \\
{\tilde L}_{+,cs2}(t) &=    \sqrt{\Gamma_{+,cs2}(t)}\, |\tilde \psi_{\uparrow +}\rangle \langle \tilde \psi_{\downarrow -}|.
\end{align}
Finally the excitaiton rates are related to the relaxation rates via the fluctuation-dissipation theorem, which for the environment at thermal equilibrium gives:
\begin{align}
    \Gamma_{+,c}(t)  &\approx \Gamma_{-,c}(t) e^{-\beta \Omega_0(t)} \nonumber \\
    \Gamma_{+,s}(t)  &\approx \Gamma_{-,s}(t) e^{-\beta E_Z} \nonumber \\
    \Gamma_{+,cs1}(t)  &\approx \Gamma_{-,cs1}(t) e^{-\beta (\Omega_0 (t)- E_Z)} \nonumber \\
    \Gamma_{+,cs2}(t)  &\approx \Gamma_{-,cs2}(t) e^{-\beta (\Omega_0(t) + E_Z)}
\end{align}
So together the AME can be characterized by four relaxation rates $\Gamma_{-,c}(t)$, $\Gamma_{-,s}(t)$, $\Gamma_{-,cs1}(t)$ and $\Gamma_{-,cs2}(t)$, that will be computed in Appendix.~\ref{app:relax}

\subsection{Equations of motion in absence of spin-orbit coupling}
We finally use Eq.~\eqref{eq:ame_app} to write equation of motion for the case where evolution is adiabatic (slow sweep rates) and the spin-orbit is not present, or negligibly small. The first conditions corresponds to avoiding dynamical term in the adiabatic Hamiltonian, such that $\tilde H \approx \tilde H_0$. The second condition, means that only charge transitions are possible (transitions c in Table~\ref{table:transitions}). As a result the equation for probability of occupying ground and excited state $\dot P_{s\pm}(t) = \langle \tilde\psi_{s\pm}(t)| \hat {\tilde \varrho}_{q,I}(t)| \tilde \psi_{s\pm}(t)\rangle$ reads:
\begin{align}
    \dot P_{s\pm}(t) &=  P_{s\mp}(t) \Gamma_{s\pm}(t)- P_{s\pm}(t) \Gamma_{s\mp}(t),
\end{align}
while the coherence between the spin-like states in the excited $W_+ = \langle \tilde\psi_{\uparrow+}(t)| \hat {\tilde \varrho}_{q,I}(t)| \tilde \psi_{\downarrow-}(t)\rangle$ and ground $W_- = \langle \tilde\psi_{\uparrow-}(t)| \hat {\tilde \varrho}_{q,I}(t)| \tilde \psi_{\downarrow-}(t)\rangle$ states reads:
\begin{equation}    
\dot W_\pm(t) =- W_\pm(t) \Gamma_\mp(t) + W_\mp(t) \Gamma_\pm(t)e^{\mp i\int_{t_0}^t \Delta \Omega(t') \text{d}t'}
\end{equation}

The above equation shows that the spin coherence can decrease due to inelastic transitions between the adiabatic states. This can be seen in the language of perturbation theory, i.e. finding the first order correction $W_-^{(1)}(t)$ to unperturbed $W_+(0) =1$, $W_-(0)=0$, which gives:
\begin{equation}
     W_-^{(1)}(t) = \int_0^t \text{d}t' \Gamma_-(t') \exp(-\int_{t_0}^{t''}\Gamma_-(t) + i \Delta \Omega(t'') ).
\end{equation}
For the constant $\Gamma_-$ and $\Delta \Omega = \Delta E_Z$ we can compute the phase error associated with orbital relaxation $Q_\varphi = (1-|W_-^{(1)}(t)|)$:
\begin{equation}
Q_\varphi = \frac{1}{2} \left(\frac{\Delta E_Z}{\Gamma_-}\right)^2,
\end{equation}
which reconstructed Eq.\eqref{eq:phi_transitions} for initially occupied excited state $P_e = 1$. In the opposite limit of relatively large $\Delta E_Z \geq \Gamma_\pm$, we would have $\dot W_\pm(t) = -W_{\pm(t)} \Gamma_{\mp}(t)$, which effectively recovers the global secular approximation, i.e. the orbital relaxation does no longer preserve any spin coherence.

\section{Relaxation rates}
\label{app:relax}
We now follow \cite{Krzywda_PRB21} and find expressions for the charge and spin relaxation rate, which will be evaluated for Si and GaAs devices in the next section \ref{app:relax_params}.

The relaxation is caused by the interaction with environment. We cast electron-environment in the form:
\begin{equation}
    \hat V = \frac{1}{2}\hat V_x \hat \tau_x + \frac{1}{2}\hat V_z \hat \tau_z,
\end{equation}
where $\tau_{i}$ are written in the dot basis $\ket{L/R}$ and the $\hat V_i$ are acting on the environmental degrees of freedom only. They correspond to noise in tunnel coupling $\hat V_x$ and noise in detuning $\hat V_z$, which are assumed to be independent, i.e. $\langle \hat V_x \hat V_z \rangle = 0$. Using Fermi Golden rule the relaxation rate between any of the states can be written as:
\begin{equation}
    \Gamma_{sq\to s'q'} = \frac{1}{4}\sum_{i=x,z} \big|\bra{\psi_{sq}} \hat \tau_i \ket{\psi_{s'q'}}\big|^2 S_i(E_{sq} - E_{s'q'}),
\end{equation}
where $S_i(\Omega)$ is the spectral density of the environment, while the time-dependence of the relaxation rate $\Gamma_{sq \to s'q'}(t)$, that results from the time-dependence of the states $\ket{\psi_{sq}(t)}$ and the energies $E_{sq}(t)$ was omitted for brevity. 

\subsection{Charge relaxation $\Gamma_{-.c}(t)$}
For the charge relaxation we consider the transition between spin-diagonal states, i.e. $\ket{\psi_{s+}} \to \ket{\psi_{s-}}$, i.e.
\begin{equation}
    \Gamma_{s+ \to s-}(t) = \frac{1}{4}\sum_{i=x,z} \big|\bra{\psi_{s+}(t)} \hat \tau_i \ket{\psi_{s-}(t)}\big|^2 S_i(\Omega_s[t]),
\end{equation}
Using definition of the instantaneous states Eq.~\eqref{eq:adiab}, we have:
 \begin{align}
 \big|\bra{\psi_{s+}(t)} \hat \tau_x \ket{\psi_{s-}(t)}\big|^2 &= \sin^2\theta_s(t) \nonumber \\
 \big|\bra{\psi_{s+}(t)} \hat \tau_z \ket{\psi_{s-}(t)}\big|^2 &= \cos^2\theta_s(t),
 \end{align}
 where $\sin\theta_s(t) = 2t_c/\Omega_s(t)$. Following assumption from \ref{app:AME}, in computing relaxation rate we neglect small difference between the orbital energies, and in this way find the spin-independent rate:
 \begin{equation}
     \Gamma_{-,c}(t) \approx \frac{1}{4}\bigg(\cos^2\theta_0[t] S_x(\Omega_0[t]) + \sin^2\theta_0[t] S_z(\Omega_0[t])\bigg),
 \end{equation}
 where $\tan\theta_0 = 2t_c/\epsilon(t)$.
 \subsection{Spin relaxation $\Gamma_s(t)$}
The spin relaxation is activated by the spin-orbit coupling, which hybridises spin-up and spin-down states around avoided crossing, i.e. the correction to states due to presence of spin orbit interaction from Eq.~\ref{eq:Vso} can be written as:
\begin{equation}
    \ket{\psi_{sq}'} = \ket{\psi_{sq}} + \sum_{q's'\neq qs} \frac{\bra{\psi_{s'q'}} \hat V_{so}\ket{\psi_{sq}}}{E_{sq} - E_{s'q'}} \ket{\psi_{q's'}}.
\end{equation}
As an illustration for the ground state we have:
\begin{align}
    \ket{\psi_{-\downarrow}'} = \ket{\psi_{-\downarrow}} &+\frac{1}{2}\,\frac{a^*\sin\tfrac{\Delta \theta}{2} - b\cos \overline \theta}{\Delta \Omega -E_Z}\ket{\psi_{-\uparrow}} \nonumber \\&- \frac{1}{2}\,
    \frac{a\cos\tfrac{\Delta \theta}{2}- b\sin\overline \theta}{\overline \Omega + E_Z}\ket{\psi_{+\uparrow}},
\end{align}
in which $\Delta \theta = \theta_\uparrow - \theta_\downarrow$, $\overline \theta = (\theta_\uparrow + \theta_\downarrow)/2$ and similarly $\Delta \Omega = \Omega_\uparrow - \Omega_\downarrow$,  $\overline \Omega = \Omega_\uparrow + \Omega_\downarrow$. For the second lowest in energy eigenstate we have:
\begin{align}
    \ket{\psi_{-\uparrow}'} = \ket{\psi_{-\uparrow}} &+ \frac{1}{2}\,\frac{a\sin\tfrac{\Delta \theta}{2} - b\cos \overline \theta}{E_Z-\Delta \Omega}\ket{\psi_{-\downarrow}} \nonumber\\&+\frac{1}{2}\, \frac{a^*\cos\tfrac{\Delta \theta}{2}+ b\sin\overline \theta}{\overline \Omega-E_Z}\ket{\psi_{+\downarrow}}.
\end{align}
The hybridised states can be substituted for the expression for spin relaxation rate, i.e.
\begin{align}
\Gamma_{\uparrow q\to\downarrow q}(t) = \frac{1}{4} \sum_i \bigg|\bra{\psi_{q\uparrow}'(t)} \hat \tau_i\ket{\psi_{q\downarrow}'(t)}\bigg|^2 S_i(E_Z +q\Delta \Omega[t]) 
\end{align}
Again following assumptions made in the derivation of AME (see Appendix~\ref{app:AME}) we use the fact that $\Delta E_Z$ is much smaller than any other energy scale, which allows us to neglect a small difference between spin relaxations in ground and excited state associated with $\Delta \Omega = \Omega_\uparrow - \Omega_\downarrow$. This ammounts to setting
$\Delta \theta\to 0$, $\Delta \Omega \to 0$, $\overline \theta \to \theta_0 = \arcsin(2t_c/\Omega_0)$, $\overline \Omega  = \Omega_0 = \sqrt{(2 t_c)^2 + \epsilon^2(t)}$ and writing:
\begin{equation}
    \Gamma_{-,s}(t) = \bigg|\frac{a E_Z  + b t_c}{\Omega_0^2(t) - E_Z^2} \bigg|^2 \Gamma_{-\text{charge}}(E_Z)
\end{equation}

 \subsection{Spin-charge relaxation $\Gamma_{sc}(t)$}
We finally compute the relaxation rate associated with simultaneous change of both orbital and spin. At finite $E_Z$ we have two distinct transitions:
\begin{align}
    \Gamma_{\uparrow +\, \to \,\downarrow -}(t) = \frac{1}{4}\sum_{i} \big|\langle\psi_{\uparrow +}'(t)| \hat \tau_i |\psi_{\downarrow -}'(t)\rangle\big|^2 S_i( \overline \Omega(t) + E_Z )\nonumber \\
     \Gamma_{\downarrow +\, \to \,\uparrow -}(t) = \frac{1}{4}\sum_{i} \big|\langle\psi_{\downarrow +}'(t)| \hat \tau_i |\psi_{\uparrow -}'(t)\rangle\big|^2 S_i( \overline \Omega(t) - E_Z ),
\end{align}
note that now since $E_Z$ is not essentially much smaller then $\overline \Omega \approx \Omega_0(t)$, the transitions should be associated with two distinct Lindbladians, meaning that no spin coherence is expected to survive the relaxation.

To compute the matrix elements we use perturbation theory (see above) and write:
\begin{align}
    \ket{\psi_{\uparrow+}'} = \ket{\psi_{\uparrow+}} &+\frac{1}{2}\,\frac{a\sin\tfrac{\Delta \theta}{2} + b\cos \overline \theta}{\Delta \Omega + E_Z}\ket{\psi_{+\downarrow}} \nonumber \\&- \frac{1}{2}\,
    \frac{a^*\cos\tfrac{\Delta \theta}{2}+ b\sin\overline \theta}{\overline \Omega + E_Z}\ket{\psi_{\downarrow-}}, \nonumber \\
    \ket{\psi_{\downarrow+}'} = \ket{\psi_{\downarrow+}} &-\frac{1}{2}\,\frac{a^*\sin\tfrac{\Delta \theta}{2} + b\cos \overline \theta}{\Delta \Omega +E_Z}\ket{\psi_{\uparrow+}} \nonumber \\&+ \frac{1}{2}\,
    \frac{a\cos\tfrac{\Delta \theta}{2}- b\sin\overline \theta}{\overline \Omega - E_Z}\ket{\psi_{\uparrow-}}.
\end{align}
Interestingly, in the limit of small $\Delta E_Z$, i.e. $\Delta \theta \to 0$, and in the leading order, the orbital relaxation with a spin-flip is driven solely by the synthetic spin-orbit interaction, i.e. 
\begin{align}
\Gamma_{-,cs1}(t) &= \bigg|\frac{E_Z b \cos\theta_0}{E_Z^2 - \Delta \Omega(t)^2}\bigg|^2 \Gamma_{-,c}(\Omega_0[t]+E_Z) \, \nonumber \\
&\approx \frac{b^2 \epsilon^2}{E_Z^2 \Omega_0^2} \Gamma_{-,c}(\Omega_0[t]+E_Z).
\end{align}
where in the approximate expression we assumed $E_Z \gg \Delta \Omega > \Delta E_Z$. Similarly we have:
\begin{equation}
    \Gamma_{-,cs2}(t) \approx  \frac{b^2 \epsilon^2}{E_Z^2 \Omega_0^2}  \Gamma_{-,c}(\Omega_0[t] - E_Z).
\end{equation}

In view of results above we now argue that charge transition with spin flip is the higher order correction to transfer error, and for this reason is negligibly small to all the other sources of errors considered in this paper. Firstly, such a charge excitation with a spin flip is highly improbable event, which can be linked to the Boltzman factor $e^{-\beta[\Omega_0(t) \pm E_Z]}$ being negligibly small everywhere apart from the vicinity of avoided crossing (see charge relaxation). At this region of small detuning $\Gamma_{+,cs} \propto \epsilon^2/\Omega_0^2$ is small as well. The same arguments holds for the relaxation, which is unlikely to happen unless $\epsilon \geq 2t_c$. Also a prefactor of $b^2/E_Z^2$ is expected to be negligibly small unless the transition takes place at manipulation region or at very low magnetic fields. As a result we conclude this process can be neglected in the analysis, which we confirm by including it in the numerical simulation of AME without a change in overall result.

\section{Relaxation rates in Si and GaAs: Spectral densities}
\label{app:relax_params}
We first consider charge relaxation caused by the charge noise. The noise in detuning is modeled by extrapolating typically measured power spectra of 1/f and Johnon noise of the form:
\begin{equation}
S_z^{1/f}(\Omega) = \frac{S_1 \omega_1}{\Omega},\,\, S_z^{J}(\Omega) = \frac{Z}{R_q}\frac{\Omega}{1-e^{-\beta\Omega}}
\end{equation}
where we use $Z = 50\Omega$ and $R_q = \pi/e^2 \approx 13k\Omega$ \cite{Huang_PRB13}.
\begin{figure}
    \centering
    \includegraphics[width=\columnwidth]{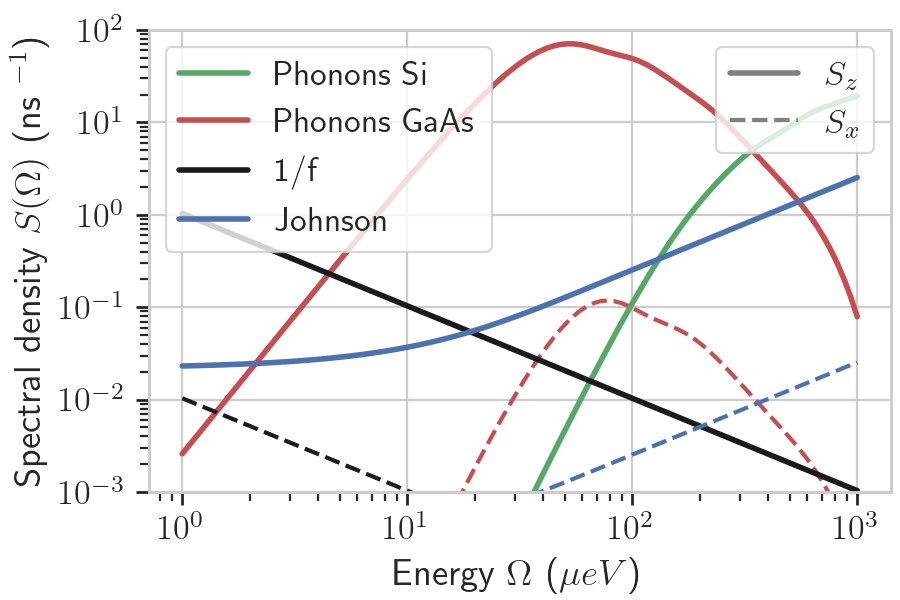}
    \caption{Spectral densities $S_{i}(\Omega)$ as a function of the energy transfer $\omega$ for a different mechanisms: Phonons in Si (green), phonons in GaAs (red), 1/f noise (black), Johnson noise (blue). With solid line we depict $S_z(\Omega)$, i.e. fast noise in detuning, while dashed line represent $S_x(\Omega)$, i.e. fast noise in tunnel coupling.}
    \label{fig:enter-label}
\end{figure}

We assume the noise in tunnel coupling is independent and its power spectrum is rescaled by a factor of $\alpha = 10^{-2}$, i.e. $S_x(\Omega) = \alpha S_z(\Omega)$. This can be related to a typically small overlap of the wavefunction and empirically observed difference between lever arms of barrier and plunger gates \cite{unseld20232d}.

\begin{figure}[h!]
    \centering
    \includegraphics[width=\columnwidth]{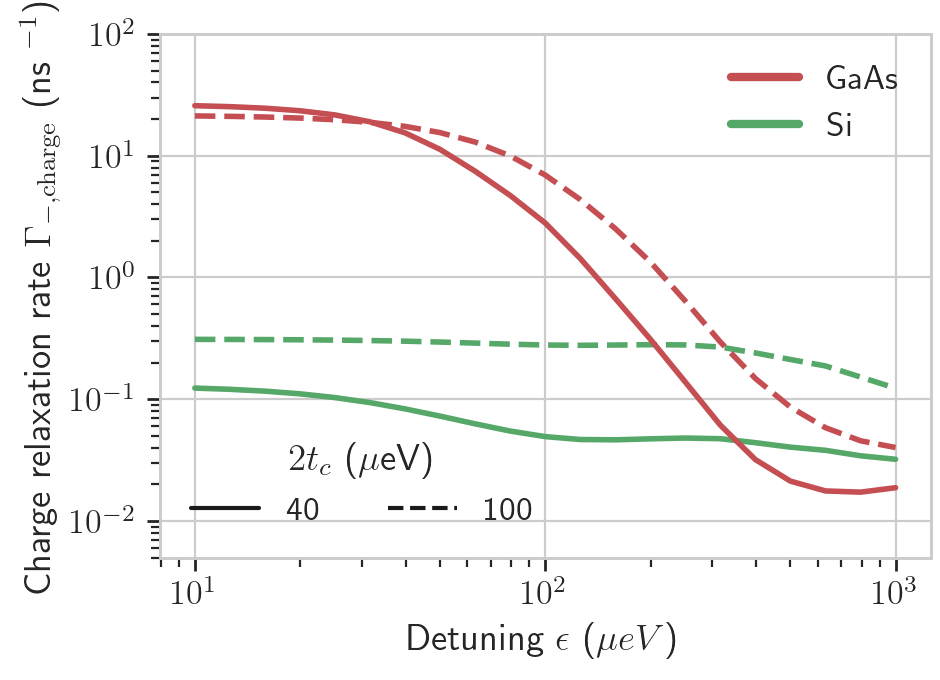}
    \caption{Charge relaxation rate in GaAs (red) and Si (green) as a function of deutning for two different tunnel couplings considered in this paper $2t_c = 40\mu$eV (solid line) and $2t_c = 100\mu$eV (dashed line). }
    \label{fig:enter-label}
\end{figure}

As a second source of the relaxation we consider phonons, for which the spectral densities can be derived from the general form of phonon-electron interaction: 
\begin{equation}
    \hat V_{\text{el}-\text{ph}}  = \sum_{j=p,d} \sum_{\\ \lambda=L,T} \sum_{\mathbf{k}} \eta_{\mathbf{k},\lambda} v_{\mathbf{k},\lambda}^{(j)} (b_{\mathbf{k},\lambda} + b_{-\mathbf{k},\lambda}^\dagger) e^{i\mathbf{k}\mathbf{r}},
\end{equation}
where we sum over piezoelectric (p) and deformation (d) phonons and their polarisation: longitudinal (L) and transverse (T), as well as the wavevector $\mathbf{k}$. The couplings are given by $\eta_{\mathbf{k},\lambda} = \sqrt{k/2\varrho c_\lambda V}$ and :
\begin{equation}
v_{\mathbf{p},\lambda}^{(p)} =\frac{\chi_p}{k}\,, \,v_{\mathbf{d},L}^{(d)} = \Xi_d + \Xi_u \frac{k_z^2}{k^2}\,,\,v_{\mathbf{d},T}^{(d)} = -\frac{k_{x,y}k_z}{k^2}
\end{equation}
Using Fermi Golden rule the Spectral density of phonons is given by:
\begin{align}
\label{eq:couplings}
    S_i(\Omega) &= 2\pi\sum_{\mathbf{k}, \lambda,j} \big|F_{i,\mathbf{k}}(\mathbf{r})v_{\mathbf{k},\lambda}^{(j)}\eta_{\mathbf{k},\lambda}  \big|^2  \delta\big(\Omega - \omega_{\mathbf{k},\lambda}\big) 
\end{align}
where $F_{x,\mathbf{k}}(\mathbf{r}) = 2\bra{L} e^{i\mathbf{k}\mathbf{r}}\ket{R}$, $F_{z,\mathbf{k}}(\mathbf{r}) = \bra{L} e^{i\mathbf{k}\mathbf{r}}\ket{L} - \bra{R} e^{i\mathbf{k}\mathbf{r}}\ket{R} $. We use linear dispersion relation $\omega_{\mathbf{k},\lambda} = c_\lambda |k|$. The spectral densities can now computed numerically by substituting \eqref{eq:hund_mulican}, substituting coupling constants \eqref{eq:couplings}, and changing sum into an integral, i.e. $\sum_{\textbf{k}} = \frac{V}{(2\pi)^3} \int \text{d}\mathbf{k}$.

\begin{figure}[h!]
    \centering
    \includegraphics[width=\columnwidth]{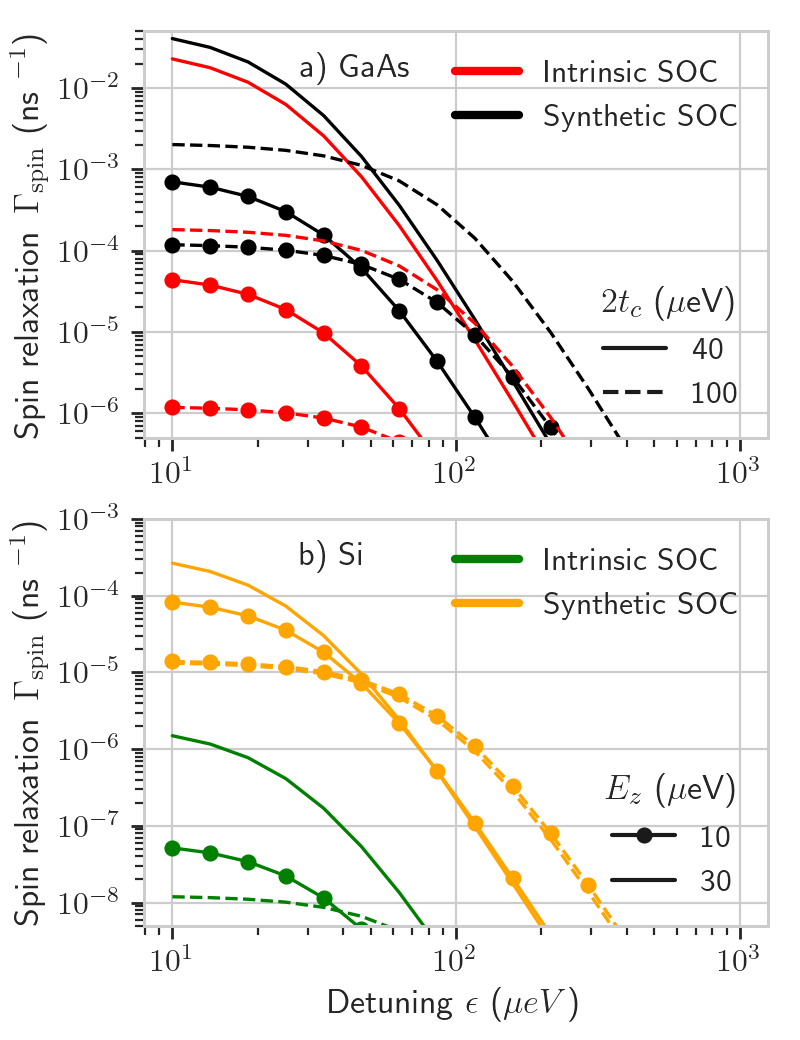}
    \caption{Spin relaxation rate in GaAs (a) and Si (b). In both panels we compare two tunnel couplings ($2t_c = 40\mu$eV (solid line), $2t_c = 100\mu$eV (dashed), as well as two spin splittings $E_Z = 30\mu$eV (line without dots), $E_Z = 10\mu$eV (lines with the dots). Finally for GaAs (a) we compare contributions from intrinsic SOC, i.e $|a| = 1\mu$eV, $b = 0$ (red) against synthetic $|a| = 0, b_\perp =1\mu$eV (black). Similarly for Si (b) we compare intrinsic SOC $|a| = 0.1\mu$eV, $b_\perp=0$ (green) against sythetic $|a| = 0, b_\perp=1\mu$eV (yellow).}
    \label{fig:enter-label}
\end{figure}

To compute the form factors $F_{i,\mathbf{k}}(\mathbf{r})$, we employ Hund-Mulliken approximation, and write hybridised states in terms of isolated Gaussian wavefunctions 
\begin{equation}
    \psi_{L_0/R_0} (\mathbf{r})= \bra{x}\ket{L_0/R_0} \psi_{yz}(y,z)= \frac{\psi_{yz}(y,z)}{(\pi L^2)^{1/4}} e^{-\frac{(x\pm d/2)^2}{2L^2}} ,
\end{equation}
parameterized by dots separation $d$ and isotropic dot size $L$. Due to finite overlap, the eigenstates are the linear combination of bare wavefunctions.
\begin{align}
\label{eq:hund_mulican}
    \ket{L} &= \ket{L_0} - \frac{\gamma}{2}\ket{R_0} \nonumber \\
    \ket{R} &= \ket{R_0} - \frac{\gamma}{2}\ket{L_0},
\end{align}
where $\gamma  \approx \bra{L_0}\ket{R_0} =  \exp(-d^2/4L^2) \ll 1$ is the overlap between the wavefunctions.

\section{Dephasing due to low-frequency noise}
\label{app:low_freq}
We provide here more detailed derivation of Eqs~\eqref{eq:deps} and \eqref{eq:t2star} , which shows that the most relevant contributions to spin-dephasing for low-frequency noise is caused by the fluctuation of detuning noise and spin-splitting noise. We start by assuming the electron stays in the two lowest-energu  instantaneous states, for which the energies are given by:
\begin{equation}
    E_{-s} = \frac{1}{2} E_Z - \frac{1}{2}\Omega_s(t)
\end{equation}
where $\Omega_s(t)  = \sqrt{(\epsilon + \sigma_s\Delta E_Z/2)^2 + (2t_c)^2}$. We define the spin splitting noise as
\begin{equation}
    \omega_\text{spin} = E_{-\uparrow} - E_{-\uparrow}
\end{equation}
and introduce the quasistatic noise, i.e. replace $\epsilon \to \epsilon + \delta \epsilon$, $t_c \to \delta t_c$, $E_Z \to E_Z + \frac{1}{2}(\delta E_{Z,l} + \delta E_{Z,r})$ and $\Delta E_Z \to \Delta E_Z + \delta E_{Z,l} - \delta E_{Z,r}$. Together the phase acquired between two spin components during a transition between $0$ and $t_f$ is given by:
\begin{equation}
    \varphi = \int_{0}^{t_f} \omega_\text{spin}(t) \text{d}t = (E_Z+\delta E_Z)t_f +  \int_{0}^{t_f} \tfrac{\Omega_{\downarrow} (t)- \Omega_{\uparrow}(t)}{2} \text{d}t,
\end{equation}
where in the first term $\delta E_Z = \tfrac{1}{2}(\delta E_{Z,L} + \delta E_{Z,R})$, which corresponds to low-frequency fluctuations in spin-splitting. 

The integrand in the second term  is related to the difference between spin-dependent orbital energies $\Delta \Omega(t) = \Omega_\downarrow(t) - \Omega_\uparrow(t)$ and can be written as:
\begin{align}
    \Delta \Omega(t) &=  \sqrt{[\epsilon + \delta \epsilon - \tfrac{1}{2}(\Delta E_Z + \delta \Delta E_Z)]^2 + (2t_c+\delta 2t_c)^2} \nonumber \\ &-  \sqrt{[\epsilon + \delta \epsilon + \tfrac{1}{2}(\Delta E_Z + \delta \Delta E_Z)]^2 + (2t_c+\delta 2t_c)^2},
\end{align}
where we denoted $\delta \Delta E_Z = \delta E_{Z,L} - \Delta E_{Z,R}$. We now exploit the fact that the noise terms are small in comparison to noiseless orbital splitting, 
\begin{equation}
    \Omega_{0,s}(t) = \sqrt{(\epsilon(t) +\tfrac{1}{2}\sigma_s \Delta E_Z)^2 + (2t_c)^2},
\end{equation}
which in linear order in the noise terms $\delta\epsilon/\Omega_{0,s}(t)$, $\delta t_c/\Omega_{0,s}(t)$ and $\delta \Delta E_Z/\Omega_{0,s}(t)$ allows to write:
\begin{align}
    \Delta \Omega(t) \approx \Delta \Omega_0(t) + \delta \epsilon\, &[\cos\theta_\downarrow(t) - \cos\theta_\uparrow(t)]\nonumber \\
    -\delta \Delta E_Z\,&[\cos\theta_\downarrow(t) + \cos\theta_\downarrow(t)]\nonumber \\+ \delta t_c\,&[\sin\theta_\downarrow(t) - \sin\theta_{\uparrow}(t)],
\end{align}
where $\Delta \Omega_0(t) = \Delta \Omega_{0,\downarrow}(t) - \Delta \Omega_{0,\uparrow}(t)$, $\cos\theta_s(t) = (\epsilon + \sigma_s \Delta E_Z/2)/\Omega_{0,s}(t)$, $\sin\theta_s(t)= 2t_c/\Omega_{0,s}(t)$. Finally we use the fact that $\Delta E_Z \ll 2t_c$, such that combination of trigonometric functions in leading order in $\Delta E_Z / 2t_c$ are given by:
\begin{align}
    \cos\theta_\downarrow(t) - \cos\theta_\uparrow(t) &\approx -\frac{\Delta E_Z}{ \Omega_{0}}\sin^2[\theta_0(t)], \nonumber \\
    \cos\theta_\downarrow(t) + \cos\theta_\uparrow(t) &\approx 2\cos[\theta_0(t)], \nonumber \\
    \sin\theta_\downarrow(t) - \sin\theta_\uparrow(t) &\approx \frac{\Delta E_Z}{2 \Omega_{0}}\sin[2\theta_0(t)],
\end{align}
where the right-hand side was expressed in terms of spin-less quantities, i.e. $\Omega_0 = \sqrt{\epsilon^2 + (2t_c)^2}$ while $\cos\theta_0 = \epsilon/\Omega_0$ and $\sin\theta_0 = 2t_c/\Omega_0$.

\begin{figure}[h!]
\includegraphics[width=\columnwidth]{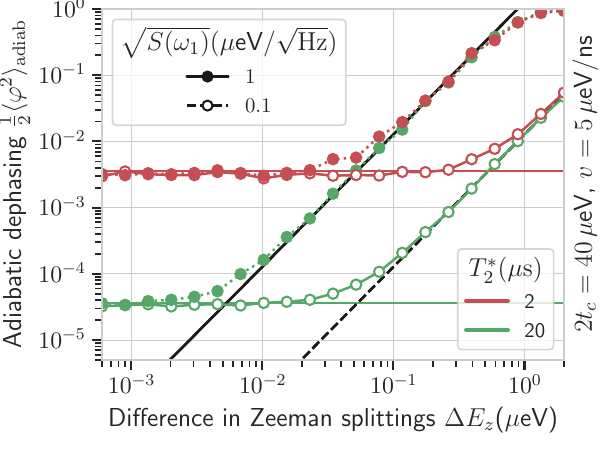}
    \caption{Dephasing during adiabatic transition, caused by the slow fluctuations of dots detuning and dot-dependent spin splittings for $2t_c = 40\mu$eV and $v = 5\mu$eV/ns. We simulate phase error by integrating the adiabatic spin splitting $E_{\uparrow-}(t) - E_{\downarrow-}(t)$ from Eq.~\eqref{eq:energy_adiab} and averaging the result over 1/f noise in detuning with $S_1 = 1 \mu$eV$^2/$Hz (solid lines, filled dots) and $S_1 = 0.1^2\mu$eV$^2/$Hz (dashed lines, hollow dots), and over spin splitting fluctuations in left and right dot modeled by two independent Ornstein-Uhlenbeck noise processes with $\sigma = \sqrt{2}/T_2^{*2}$ and correlation time $\tau_c = 1$s for two values of $T_2^* = 2\mu$s (red) and $T_2^* = 20\mu$s (green). We compare numerical results (dots) against analytical quasistatic predictions (lines) given by Eq.~\eqref{eq:deps} and Eq.~\eqref{eq:t2star}.}
    \label{fig:slow_noise}
\end{figure} 

We highlight that for the symmetric sweep, the quasistatic contribution from $\delta \Delta E_Z$ and $\delta t_c$ vanishes, and we are left with :
\begin{equation}
\label{eq:varphi_app}
    \varphi = \varphi_0 + \delta \varphi_{E_Z} + \delta \varphi_\epsilon,
\end{equation}
where $\varphi_0 = \int_{0}^{t_f}  E_Z + \Delta \Omega_0(t) \text{d}t$ is the noiseless phase difference, $\delta \varphi_{E_Z} =( \delta E_{Z,L} + \delta E_{Z,R})t_f/2$ and
\begin{equation}
    \delta \varphi_\epsilon = 2t_c\Delta E_Z \delta \epsilon \, \int_{0}^{t_f} \frac{\text{d}t}{\sqrt{\epsilon(t)^2 + (2t_c)^2}} = \frac{\Delta E_Z t_f\, \delta \epsilon}{\sqrt{v^2 t_f^2 + (2t_c)^2}},
\end{equation}
which in the typical limit of $vt_f \gg 2t_c$ reduces to:
\begin{equation}
    \delta \varphi_\epsilon \approx \frac{\Delta E_Z }{v}\, \delta \epsilon,
\end{equation}
such that for sufficiently long and symmetric sweep of detuning the small contribution to dephasing due to low-frequency noise can be written as:
\begin{equation}
    \langle \delta \varphi^2 \rangle = \langle (\delta \varphi_{E_Z} + \delta \varphi_{\epsilon})^2 \rangle = \left(\frac{t_f}{T_2^*}\right)^2 + \left(\frac{\Delta E_Z}{v}\right)^2 \sigma_\epsilon^2,
\end{equation}
where we assumed that $\langle \delta E_Z \delta \epsilon \rangle = 0 $, and the spin splitting fluctuations in the dots have the same amplitude but are statistically independent $\sqrt{2}/T_2^* = \sigma_{E_{Z,L}} = \sigma_{E_{Z,R}}$.

We perform numerical test of
an adiabatic dephasing. To do so we average $\varphi$, given by Eq.~\eqref{eq:varphi_app}, over $\delta \epsilon$, $\delta t_c$, $\delta E_{Z,L}$, $\delta E_{Z,R}$. The results are depicted in Fig.~\ref{fig:slow_noise}, where numerically simulated $\langle \delta \varphi^2 \rangle$ (dots) is shown as a function of noiseless difference in spin splittings $\Delta E_Z$ between the two dots. 

We plot the results for two different values of $T_2^* = 2,20\,\mu$s and two different amplitude of detuning noise $\langle \delta \epsilon^2 \rangle = 5,0.5\,\mu$eV, which would correspond to a 1/f noise in detuning with a typically measured spectral density at 1Hz given by $\sqrt{S(1\text{Hz})} \approx 0.5,0.05\, \mu$eV/Hz.  As it can be seen from the Figure, the numerical averages (dots) follows quasistatic predictions (black solid and dashed lines) for sufficiently large $\Delta E_Z$, for which the charge noise contribution dominates over fluctuations of Zeeman splittings. By simultaneous averaging over spin splitting fluctuations and detuning noise (previous section), we also showed that in the relevant error range (small $Q$), the charge noise and nuclear noise contribution to the transfer error $Q$ can be treated additively.

\bibliography{literature,refs_Si,refs_spin_qubits,refs_decoherence,refs_Ge}

\end{document}